\documentclass[12pt]{article}
\usepackage{amssymb}
\usepackage{graphicx}
\usepackage{amsmath}

\input{tcilatex}

\begin{document}

\begin{center}
{\Large Easy-to-Implement Two-Way Effect Decomposition}

{\Large for Any Outcome Variable with Endogenous Mediator\medskip }

\begin{tabular}{l}
Bora Kim \\ 
Department of Finance, \\ 
\ \ \ Accounting \& Economics \\ 
University Nottingham Ningbo \\ 
\ \ \ China, Ningbo 315100, China \\ 
Bora.Kim@nottingham.edu.cn%
\end{tabular}
\begin{tabular}{l}
Myoung-jae Lee* \\ 
Department of Economics, Korea University \\ 
\ \ \ Seoul 02841, Korea,\ myoungjae@korea.ac.kr; \\ 
Department of Finance, Accounting \& Economics \\ 
\ \ \ University of Nottingham Ningbo China \\ 
Ningbo 315100, China%
\end{tabular}%
\bigskip \bigskip \bigskip \bigskip 
\end{center}

Given a binary treatment $D$ and a binary mediator $M$, mediation analysis
decomposes the total effect of $D$ on an outcome $Y$ into the direct and
indirect effects. Typically, both $D$ and $M$ are assumed to be exogenous,
but this paper allows $M$ to be endogenous while maintaining the exogeneity
of $D$, which holds certainly if $D$ is randomized. The endogeneity problem
of $M$ is then overcome using a binary instrumental variable $Z$. We derive
a nonparametric \textquotedblleft causal reduced form
(CRF)\textquotedblright\ for $Y$ with either $(D,Z,DZ)$ or $(D,M,DZ)$ as the
regressors. The CRF enables estimating the direct and indirect effects
easily with ordinary least squares or instrumental variable estimator,
instead of matching or inverse probability weighting that have difficulties
in finding the asymptotic distribution or in dealing with near-zero
denominators. Not just this ease in implementation, our approach is
applicable to any $Y$ (binary, count, continuous, etc.). Simulation and
empirical studies illustrate our approach.\bigskip \bigskip \bigskip
\bigskip 

* Corresponding author.

\textbf{Running Head}: effect decomposition with endogenous mediator.


\textbf{Statements/declarations on no competing interests, compliance with
ethical standards, and no AI\ usage}:\ No human/animal subject is involved
in this research, and there is no conflict of interest to disclose. Also, no
generative AI-related technique has been used for this paper.\pagebreak

\section{Introduction}

\qquad Given a binary treatment $D$, a binary mediator $M$ and an outcome
variable $Y$, researchers are often interested in the direct effect of $D$
on $Y$ and the indirect effect of $D$ on $Y$ through $M$. The total effect
is then the sum of the direct and indirect effects. This is an important
issue in many disciplines of science; see reviews in MacKinnon et al.
(2007), Pearl (2009), Imai et al. (2010), TenHave and Joffe (2012), Preacher
(2015), VanderWeele (2015), Nguyen et al. (2021), and Lee (2024), among
others.

\qquad Typically, both $D$ and $M$ are assumed to be exogenous (e.g., Huber
et al. 2018; Bellani and Bia 2019, among many others), but we allow for $M$,
not $D$, to be endogenous with a binary instrumental variable (IV) $Z$ for $%
M $ available. An empirical example is Chen et al. (2019): effects of having
a brother ($D$) on high-school-completion/college-entry of the firstborn,
where $M$ is the number of siblings greater than two or not. The direct
effect is $D$ negatively affecting $Y$ (sibling rivalry), and the indirect
effect is through a smaller $M$ due to strong son-preference; having twins
at the second birth is $Z$.

\qquad As for the literature on allowing for endogeneity of either $D$ or $M$%
, Imai et al. (2013) allowed for endogenous $M$, but their $M$ should be
partly controllable, which is not necessary in our approach. Mattei and
Mealli (2011) allowed for endogenous $M$ when $D$ is randomized to propose a
bounding approach, whereas our approach does not require a randomized $D$.
Joffe et al. (2008) allowed for both $D$ and $M$ to be endogenous when only
a single IV is available under linear model assumptions, while ruling out
interaction terms $DM$ and $ZD$ that can appear freely in our nonparametric
approach.

\qquad Burgess et al. (2015) also allowed for both $D$ and $M$ to be
endogenous while ruling out the interaction $DM$ and effect heterogeneity,
but their framework is parametric whereas ours is nonparametric. Fr\"{o}lich
and Huber (2017) further allowed for both $D$ and $M$ to be endogenous with
a binary IV for $D$ and a discrete/continuous IV for a discrete/continuous $M
$; their approach is nonparametric, decomposing the total effect on
\textquotedblleft the IV compliers\textquotedblright\ into direct and
indirect effects, whereas our effect decomposition using \textquotedblleft
mediator principal stratification\textquotedblright\ is not for the IV
compliers. Rudolph et al. (2024) study settings with endogenous treatment
and endogenous mediator using two instruments. Their analysis, however,
focuses on interventional (in)direct effects rather than natural ones,
unlike our paper. As Miles (2023) showed, interventional indirect effects
may be nonzero even when all individual-level indirect effects are
zero.\bigskip 

\qquad\ Before examining endogenous $M$, we now review some findings for
exogenous $(D,M)$ that this paper aims to generalize. With $(D,M)$
exogenous, consider two potential versions $M^{d}$ of $M$ corresponding to $%
D=0,1$, and the four potential outcomes $Y^{dm}$ for $D=0,1$ and $M=0,1$.
Also, define the potential outcome \textquotedblleft when $M$ is allowed to
take its natural course given $D=d$\textquotedblright :%
\begin{equation*}
Y_{d}\equiv Y^{d,M^{d}}.
\end{equation*}%
Then the mean total effect of $D$ is $%
E(Y_{1}-Y_{0})=E(Y^{1,M^{1}}-Y^{0,M^{0}})$, which can be estimated with
matching, regression adjustment, inverse probability weighting, etc.; see,
e.g., Lee and Lee (2022) and Choi and Lee (2023a) for reviews on treatment
effect estimators. The question is how to decompose the total effect into
sub-effects of interest.

\qquad The well-known two-way decompositions (Pearl 2001; Robins 2003) are:%
\begin{eqnarray}
(a) &:&E(Y^{1,M^{1}}-Y^{1,M^{0}})+E\{Y^{1,M^{0}}-Y^{0,M^{0}}\};  \notag \\
(b) &:&E\{Y^{1,M^{1}}-Y^{0,M^{1}}\}+E(Y^{0,M^{1}}-Y^{0,M^{0}}).  \TCItag{1.1}
\end{eqnarray}%
These two decompositions differ only in which variable is subtracted and
added: $Y^{1,M^{0}}$ in (a), and $Y^{0,M^{1}}$ in (b). Going further from
the two-way decompositions, VanderWeele (2013) proposed a three-way
decomposition, and VanderWeele (2014) proposed a four-way decomposition that
includes the other existing decompositions as special cases.

\qquad With many decompositions of the total effect available, it is not
clear which one to use. Recently, Lee (2024) advocated a particular
three-way decomposition based on a \textquotedblleft mediative principal
stratification\textquotedblright :%
\begin{eqnarray}
&&\ E(Y^{10}-Y^{00})+E\{(Y^{01}-Y^{00})(M^{1}-M^{0})\}+E(\Delta Y^{\pm
}M^{1})\ \ \ \ \ \text{where}  \TCItag{1.2} \\
&&\ \Delta Y^{\pm }\equiv
Y^{11}-Y^{01}-Y^{10}+Y^{00}=Y^{11}-Y^{00}-(Y^{01}-Y^{00})-(Y^{10}-Y^{00}). 
\notag
\end{eqnarray}%
This appeared also in VanderWeele (2014) with different notations. Lee
(2024) then showed how to identify and estimate the three sub-effects in
(1.2).

\qquad In (1.2), the first term $E(Y^{10}-Y^{00})$ is the \textit{direct
effect}, as the $d$ in $Y^{d0}$ changes from $0$ to $1$. The second term is
the \textit{indirect effect}, as the $d$ in $M^{d}$ changes and then the $m$
in $Y^{0m}$ changes. The third term is the \textit{interaction effect}
(i.e., the effect of $DM$), because the `net effect' of $DM$ is $\Delta
Y^{\pm }$ which is the `gross effect' $Y^{11}-Y^{00}$ of $DM$ minus the
`partial effects' $Y^{01}-Y^{00}$ of $M$ and $Y^{10}-Y^{00}$ of $D$ (Choi
and Lee 2018).

\qquad Surprisingly, Lee (2024) showed that (1.1)(b) is the same as (1.2)
when the interaction effect is regarded as part of the direct effect, as the
direct effect can vary depending on the level of $M$. That is, the first
part of (1.1)(b) is the sum of the first and third terms in (1.2), and the
second part of (1.1)(b) is the middle term in (1.2). Thus, this finding
answers the big question \textquotedblleft which is preferred in
(1.1)?\textquotedblright : (1.1)(b) is preferred.\bigskip

\qquad Turning back to exogenous $D$ and endogenous $M$ with a binary IV $Z$%
, since $Z$ should affect $M$, the double-indexed $M^{dz}$ instead of $M^{d}$
is the potential version of $M$ corresponding to $D=0,1$ and $Z=0,1$; $%
Y^{dm} $ is still valid, as the IV $Z$ does not affect $Y$ directly. There
are two possibilities to generalize (1.2) when $M^{dz}$ appears:%
\begin{eqnarray}
z &=&0:E(Y^{10}-Y^{00})+E\{(Y^{01}-Y^{00})(M^{10}-M^{00})\}+E(\Delta Y^{\pm
}M^{10}),  \notag \\
z &=&1:E(Y^{10}-Y^{00})+E\{(Y^{01}-Y^{00})(M^{11}-M^{01})\}+E(\Delta Y^{\pm
}M^{11}).  \TCItag{1.3}
\end{eqnarray}%
The former with $z=0$ (i.e., no IV) may look like the right generalization
of (1.2), but it differs from the total effect $E(Y|D=1)-E(Y|D=0)$ when $D$
is randomized, as to be seen below. Thus \textit{we take the }$Z$\textit{%
-weighted average of the two expressions in (1.3) as the desired
decomposition}, which equals $E(Y|D=1)-E(Y|D=0)$ for a randomized $D$.

\qquad Differently from Lee (2024) for exogenous $M$, however, identifying
and estimating the three sub-effects turns out to require implausible
assumptions. Hence, we merge the interaction effect into the direct effect.
With `M2M' standing for `Main 2-way Mediator-based decomposition', our
target is the following 2-way decomposition based on (1.3):%
\begin{eqnarray}
&&\ E[\ \{Y^{10}-Y^{00}+(Y^{01}-Y^{00})(M^{10}-M^{00})+\Delta Y^{\pm
}M^{10}\}\cdot (1-Z)  \notag \\
&&\ \ \ \ +\{Y^{10}-Y^{00}+(Y^{01}-Y^{00})(M^{11}-M^{01})+\Delta Y^{\pm
}M^{11}\}\cdot Z\ ]  \TCItag{M2M} \\
&&\ =E[\ Y^{10}-Y^{00}+(Y^{01}-Y^{00})(M^{10}-M^{00})+\Delta Y^{\pm }M^{10} 
\notag \\
&&\ \ \ \ \ \ \ \ +\{\Delta Y^{\pm }(M^{11}-M^{10})\ +\
(Y^{01}-Y^{00})\Delta M^{\pm }\}\cdot Z\ ];  \TCItag{1.4}
\end{eqnarray}%
$\Delta M^{\pm }$ is defined analogously to $\Delta Y^{\pm }$, and (1.4)
holds by collecting the terms with $\pm Z$.

\qquad A structural form (SF) has parameters governing the behavior of the
subject, so that they are causal parameters of interest. In contrast, a
reduced form (RF) is derived from multiple SF's. Since RF parameters are
derived from SF parameters, they are not of interest per se. For M2M, this
paper uses \textquotedblleft causal reduced forms (CRF's)\textquotedblright\
for $M$ and $Y$, which fall in between SF and RF, as CRF's are RF's but with
causal parameters.

\qquad As will be seen below, our CRF's for $M$ and $Y$ are nonparametric
with $(D,Z,DZ)$ or $(D,M,DZ)$ on the right-hand side, and slopes of these
are $X$-conditional effects where $X$ is exogenous observed covariates.
E.g., the slope of $D$ in a CRF for $Y$ is the $X$-conditional total effect
of $D$ with $z=0$ in (1.3). We approximate those unknown $X$%
-functions/slopes/effects linearly, and estimate them with ordinary least
squares (OLS) or instrumental variable estimator (IVE), which makes our
approach much easier to implement than other estimators in the literature.
Since the $X$-conditional effects such as $E(Y^{10}-Y^{00}|X)$ are of RF
variety, specifying them is in general less riskier than specifying SF's
with constant effect parameters (that is usually done in practice).

\qquad In the remainder of this paper, Section 2 derives the M- and Y-CRF's
that hold for any $Y$ (binary, count, continuous, etc.), which then lead to
M2M. Section 3 estimates the direct and indirect effects with OLS and IVE.
Section 4 and 5 present simulation and empirical studies. Finally, Section 6
concludes this paper. We consider independent and identically distributed
observations across $i=1,...,N$ units, and as has been done already, the
subscript $i$ as in $Y_{i}$ will be often omitted.

\section{\textbf{Causal Reduced Forms (CRF's)}}

\qquad In this section, first, we list our five main assumptions. Then we
derive a M-CRF and two Y-CRF's. In the process, we show how the direct and
indirect effects in M2M can be estimated, with our estimators presented in
the next section.

\subsection{Five Main Assumptions}

\qquad With `$\amalg $' standing for independence, our first three main
assumptions are:%
\begin{eqnarray*}
&&\text{\textbf{C(a)} `}D,Z\text{ exogeneity'}:(D,Z)\amalg (Y^{dm},M^{dz},\
d,m,z=0,1)|X\text{ for all }X; \\
&&\text{\textbf{C(b) }`}M^{dz}\text{ monotonicity'}:M^{dz}\leq M^{d^{\prime
}z^{\prime }}\text{ \ \ for any }d\leq d^{\prime}\text{ and}\ z\leq
z^{\prime }; \\
&&\text{\textbf{C(c)}\ `Support overlap'}:0<P(D=d,M=m,Z=z|X)\text{ for all }X%
\text{, \ }d,m,z=0,1.
\end{eqnarray*}

\qquad C(a) is that $D$ and $Z$ are exogenous: given $X$, $(D,Z)$ are
independent of all potential variables. C(b) is a monotonicity condition on $%
M^{dz}$ analogous to that in Imbens and Angrist (1994) for a single-indexed $%
D^{z}$, when $D$ is endogenous with a binary IV $Z$ but without any
mediator. Nevertheless, since $M^{dz}$ is double-indexed, C(b) differs much
from the monotonicity condition for $D^{z}$; monotonicity with a
double-index and the ensuing complications relative to single-indexed cases
can be seen in Choi and Lee (2023b) and references therein. C(c) is the
usual support overlap condition to ensure the existence of all relevant
subpopulations defined by $(D,M,Z)$.

\qquad To introduce our fourth and fifth assumptions, define
\textquotedblleft mediative compliers (CP's)\textquotedblright\ as those who
change their $M$ in reaction to a $D$ or $Z$ change. We consider three types:%
\begin{equation*}
\text{IV-CP:}\ M^{01}=1,\ M^{00}=0;\text{ \ \ \ \ TR}_{z}\text{-CP:}\
M^{1z}=1,\ M^{0z}=0
\end{equation*}%
where IV-CP stands for \textquotedblleft instrument CP\textquotedblright ,
and TR$_{z}$ stands for \textquotedblleft treatment CP with $Z=z$%
\textquotedblright . It is possible for a subject to be multiple types of
CP's. E.g., consider the monotonicity-respecting subject $%
(M^{00},M^{01},M^{10},M^{11})=(0,1,1,1)$, who is an IV-CP and TR$_{0}$-CP,
but not TR$_{1}$-CP. In contrast, another monotonicity-respecting subject $%
(M^{00},M^{01},M^{10},M^{11})=(0,0,0,1)$ is a TR$_{1}$-CP, but neither IV-CP
nor TR$_{0}$-CP.

\qquad Our fourth and fifth main assumptions are:%
\begin{eqnarray*}
\text{\textbf{C(d) }`Equal IV }M\text{-effects'} &:&E(Y^{01}-Y^{00}|\text{TR}%
_{0}\text{-CP},X)=E(Y^{01}-Y^{00}|\text{IV-CP},X); \\
\text{\textbf{C(e) }`Equal TR }M\text{-effects'} &:&E(Y^{01}-Y^{00}|\text{TR}%
_{0}\text{-CP},X)=E(Y^{01}-Y^{00}|\text{TR}_{1}\text{-CP},X).
\end{eqnarray*}%
C(d) is that $M^{10}-M^{00}=1$ can be replaced with $M^{01}-M^{00}=1$ in the
conditioning set, whereas C(e) is that $M^{10}-M^{00}=1$ can be replaced
with $M^{11}-M^{01}=1$. C(d) is critical in dealing with endogenous $M$ with
an IV, because, although we desire the indirect effect with $M$ changing due
to the $D$ change, what we can use is only the indirect effect with $M$
changing due to the $Z$ change. Hence, C(d) is likely to be essential in any
IV-based approach. C(e) is a kind of \textquotedblleft IV
irrelevance\textquotedblright\ assumption, because the difference between TR$%
_{0}$ and TR$_{1}$ is the assigned value $z$ to $M^{dz}$ being $0$ versus $1$%
.

\qquad The simplest case for C(d) and C(e) to hold is $Y^{01}-Y^{00}$\ being
a constant for all subjects, which seems why C(d) and C(e) are not seen in
the literature specifying constant-effect SF's. This illustrates the\textit{%
\ hazard of using a tightly specified model: restrictions such as C(d) and
C(e) can go unnoticed, as they are easily satisfied by constant-effect SF's. 
}The appendix presents a \textquotedblleft random-effect\textquotedblright\
case for C(d) and C(e) to hold.

\subsection{CRF for M}

\qquad Recalling $\Delta M^{\pm }\equiv M^{11}-M^{01}-M^{10}+M^{00}$, as
both $D$ and $Z$ affect $M$, we have%
\begin{eqnarray}
&&\ M=(1-D)(1-Z)M^{00}+(1-D)ZM^{01}+D(1-Z)M^{10}+DZM^{11}  \notag \\
&&\ \ \ \ =M^{00}+(M^{10}-M^{00})\cdot D+(M^{01}-M^{00})\cdot Z+\Delta
M^{\pm }\cdot DZ\text{.}  \TCItag{2.1}
\end{eqnarray}%
Take $E(\cdot |D,Z,X)$ on this $M$ equation: due to C(a),%
\begin{eqnarray*}
&&E(M|D,Z,X)=\alpha _{0}(X)+\alpha _{d}(X)D+\alpha _{z}(X)Z+\alpha
_{dz}(X)DZ,\ \ \ \alpha _{0}(X)\equiv E(M^{00}|X), \\
&&\alpha _{d}(X)\equiv E(M^{10}-M^{00}|X),\ \ \alpha _{z}(X)\equiv
E(M^{01}-M^{00}|X),\ \ \alpha _{dz}(X)\equiv E(\Delta M^{\pm }|X).
\end{eqnarray*}%
Then, defining $U_{0}\equiv M-E(M|D,Z,X)$ renders Theorem 1.\bigskip

\textbf{THEOREM 1.} \textit{Under C(a), a nonparametric M-CRF holds:}%
\begin{equation}
M=\alpha _{0}(X)+\alpha _{d}(X)D+\alpha _{z}(X)Z+\alpha _{dz}(X)DZ+U_{0}%
\text{,}\ \ E(U_{0}|D,Z,X)=0,  \tag{M-CRF}
\end{equation}%
\textit{and, under C(b) and C(c), }$\alpha _{d}(X)=P($TR$_{0}$-CP$|X)>0$%
\textit{\ and} $\alpha _{z}(X)=P($IV-CP$|X)>0.$\bigskip

\textbf{Proof}: M-CRF was proven already, and observe, due to C(b) and C(c):%
\begin{eqnarray*}
&&\ \alpha _{d}(X)=E(M^{10}-M^{00}|X)=P(M^{10}=1|X)-P(M^{00}=1|X) \\
&&\ =\{P(M^{00}=0,M^{10}=1|X)+P(M^{00}=1,M^{10}=1|X)\}-P(M^{00}=1|X) \\
&&\ =\{P(M^{00}=0,M^{10}=1|X)+P(M^{00}=1|X)\}-P(M^{00}=1|X) \\
&&\ =P(M^{00}=0,M^{10}=1|X)=P(\text{TR}_{0}\text{-CP}|X)>0.
\end{eqnarray*}%
Doing analogously,%
\begin{equation*}
\alpha _{z}(X)=E(M^{01}-M^{00}|X)=P(M^{00}=0,M^{01}=1|X)=P(\text{IV-CP}%
|X)>0.\blacksquare
\end{equation*}

\qquad The M-CRF holds for any $M$ (binary, count, continuous, etc.),
although we assume binary $M$ for effect decomposition. The M-CRF is
nonparametric, as no parametric assumption was invoked, and it can be
estimated with OLS if the $\alpha $ functions are specified (e.g., linearly)
as in our empirical section. In the M-CRF, the effect of $D$ on $M$ is $%
\alpha _{d}(X)\equiv E(M^{10}-M^{00}|X)$ if $Z=0$, and $\alpha
_{d}(X)+\alpha _{dz}(X)=E(M^{11}-M^{01}|X)$ if $Z=1$. Hence, the $(X,Z)$%
-conditional effect of $D$ on $M$ is%
\begin{equation}
\alpha _{d}(X)+\alpha _{dz}(X)Z.  \tag{2.2}
\end{equation}

\qquad The term `CRF' may sound strange, but CRF has been fruitfully used in
Lee (2018, 2021), Mao and Li (2020), Choi et al. (2023), Lee and Han (2024),
Lee et al. (2023), Kim and Lee (2024), Lee et al. (2025), and Kim (2025). In
fact, a CRF with an effect constancy restriction appeared much earlier in
Angrist (2001; equations 17 and 18).

\subsection{First CRF for Y with Regressors (D,Z,DZ)}

\qquad This subsection presents a Y-CRF with $(D,Z,DZ)$ as the regressors,
whose $D$- and $DZ$-slopes render the total effect in M2M, whereas the next
subsection presents another Y-CRF with $(D,M,DZ)$ as the regressors, whose $D
$- and $DZ$-slopes renders the direct effect in M2M. Then, the indirect
effect can be found by subtracting this direct effect from the total effect.
The proofs for the two Y-CRF's are in the appendix.\bigskip

\textbf{THEOREM 2.} \textit{Under C(a) to C(c), a nonparametric Y-CRF with
the regressors }$(D,Z,DZ)$\textit{\ holds for any form of }$Y$ \textit{%
(binary, count, continuous, ...):}%
\begin{eqnarray}
&&Y=\beta _{0}(X)+\beta _{d}(X)D+\beta _{z}(X)Z+\beta _{dz}(X)DZ+U_{1},\ \ \
U_{1}\equiv Y-E(Y|D,Z,X)  \notag \\
&&\ \ \ =\beta _{0}(X)+\beta _{z}(X)Z+\{\beta _{d}(X)+\beta _{dz}(X)Z\}\cdot
D+U_{1},  \TCItag{Y-CRF1} \\
&&\beta _{0}(X)\equiv E\{Y^{00}+(Y^{01}-Y^{00})M^{00}|X\},\ \ \beta
_{z}(X)\equiv E\{(Y^{01}-Y^{00})(M^{01}-M^{00})|X\},  \notag \\
&&\beta _{d}(X)\equiv E\{Y^{10}-Y^{00}+(Y^{01}-Y^{00})(M^{10}-M^{00})+\Delta
Y^{\pm }M^{10}|X\},  \notag \\
&&\beta _{dz}(X)\equiv E\{\Delta Y^{\pm
}(M^{11}-M^{10})+(Y^{01}-Y^{00})\Delta M^{\pm }|X\}.  \notag
\end{eqnarray}%
\textit{Analogously to (2.2), }$\beta _{d}(X)+\beta _{dz}(X)Z$\textit{\ is
the total effect of }$D$ \textit{given }$(X,Z)$\textit{, which becomes the
marginal total effect (1.4) when }$(X,Z)$\textit{\ is integrated out.}%
\bigskip

\qquad Just as M-CRF is nonparametric, Y-CRF1 is also nonparametric because
no parametric assumption is invoked to derive Y-CRF1. Since $%
E(U_{1}|D,Z,X)=0 $ by construction, we can apply OLS to Y-CRF1 once all $%
\beta (X)$ functions are specified (e.g., linearly).

\qquad There appear two different indirect effects in $\beta _{d}(X)$ and $%
\beta _{z}(X)$ of Y-CRF1:%
\begin{equation}
E\{(Y^{01}-Y^{00})(M^{10}-M^{00})|X\}\text{\ \ \ and \ \ }%
E\{(Y^{01}-Y^{00})(M^{01}-M^{00})|X\};  \tag{2.3}
\end{equation}%
both are \textquotedblleft endogenous-$M$ generalizations\textquotedblright\
of $E\{(Y^{01}-Y^{00})(M^{1}-M^{0})|X\}$ in (1.2) that is for exogenous $M$. 
\textit{The former in (2.3) is the indirect effect of }$D$\textit{, which is
of interest, but the latter in (2.3) is the indirect effect of }$Z$\textit{,
which is not of interest.}

\qquad The slope $\beta _{d}(X)+\beta _{dz}(X)Z$ of $D$ is $\beta
_{d}(X)+\beta _{dz}(X)$ with $Z=1$, where two terms with $\Delta Y^{\pm }$
appear: $\Delta Y^{\pm }M^{10}$ and $\Delta Y^{\pm }(M^{11}-M^{10})$, whose
sum is just $\Delta Y^{\pm }M^{11}$. Thus,%
\begin{eqnarray}
&&\ \beta _{d}(X)+\beta _{dz}(X)  \notag \\
&&\ =E\{Y^{10}-Y^{00}+(Y^{01}-Y^{00})(M^{10}-M^{00})+(Y^{01}-Y^{00})\Delta
M^{\pm }+\Delta Y^{\pm }M^{11}|X\}  \notag \\
&&\ =E\{Y^{10}-Y^{00}+(Y^{01}-Y^{00})(M^{11}-M^{01})+\Delta Y^{\pm
}M^{11}|X\}.  \TCItag{2.4}
\end{eqnarray}%
This differs from $\beta _{d}(X)$ only in that $0$ in $M^{d0}$ is replaced
by $1$. The $(X,Z)$-conditional total effect of $D$ on $Y$ is $\beta
_{d}(X)+\beta _{dz}(X)Z$, and \textit{the marginal total effect is} $%
E\{\beta _{d}(X)+\beta _{dz}(X)Z\}$. \textit{If }$D$\textit{\ is randomized, 
}$E(Y|D=1)-E(Y|D=0)$\textit{\ can be used as the marginal total effect,
which is also }$E\{\beta _{d}(X)+\beta _{dz}(X)Z\}$\textit{\ from Y-CRF1.}

\subsection{Second CRF for Y with Regressors (D,M,DZ)}

\qquad Turning to the second Y-CRF, recall the monotonicity C(b), and define 
$\beta _{m}(X)$:%
\begin{eqnarray}
&&\ \beta _{m}(X)\equiv E(Y^{01}-Y^{00}|M^{10}-M^{00}=1,X)  \notag \\
&&\ \Longrightarrow \ \beta _{m}(X)\alpha
_{d}(X)=E(Y^{01}-Y^{00}|M^{10}-M^{00}=1,X)\cdot E(M^{10}-M^{00}|X)  \notag \\
&&\ \ \ \ \ \ \ \ \ \ \ \ \ \ \ \ \ \ \ \ \ \ \ \ \ \
=E\{(Y^{01}-Y^{00})(M^{10}-M^{00})|X\};  \TCItag{2.5}
\end{eqnarray}%
$\beta _{m}(X)$ is the effect of $M$\ for the TR$_{0}$-CP's, and $\beta
_{m}(X)\alpha _{d}(X)$\textit{\ is the indirect effect of }$D$\textit{\ with 
}$z=0$. Using this, the appendix proves Theorem 3 next.\bigskip

\textbf{THEOREM 3.} \textit{Under C(a) to C(e), a nonparametric Y-CRF with
the regressors }$(D,M,DZ)$\textit{\ holds for any form of }$Y$ \textit{%
(binary, count, continuous, ...):}%
\begin{eqnarray}
&&Y=\beta _{0}(X)-\beta _{m}(X)\alpha _{0}(X)\ +\beta _{m}(X)M\ +\ [\beta
_{d}(X)-\beta _{m}(X)\alpha _{d}(X)  \notag \\
&&\ \ +\{\beta _{dz}(X)-\beta _{m}(X)\alpha _{dz}(X)\}Z]\cdot D+U_{2},\ \ \
E(U_{2}|D,Z,X)=0  \TCItag{Y-CRF2}
\end{eqnarray}%
\textit{and} $U_{2}\equiv -\beta _{m}(X)U_{0}+Y-E(Y|D,Z,X)$ \textit{with} $%
U_{0}\equiv M-E(M|D,Z,X)$. \textit{The slope of }$D$ \textit{with }$Z$%
\textit{\ is the direct effect of }$D$\textit{\ on }$Y$\textit{\ given }$%
(X,Z)$\textit{, and thus the marginal direct effect of }$D$\textit{\ on }$Y$%
\textit{\ is }$E[\beta _{d}(X)-\beta _{m}(X)\alpha _{d}(X)+\{\beta
_{dz}(X)-\beta _{m}(X)\alpha _{dz}(X)\}Z]$\textit{.}\bigskip

\qquad Y-CRF2 is nonparametric just as M-CRF and Y-CRF1 are. $%
E(U_{2}|D,Z,X)=0$ holds because $U_{2}$ consists of two error terms with
zero $(D,Z,X)$-conditional means. Hence, IVE can be applied to Y-CRF2 with
the regressors $(1,D,M,DZ)$ and the IV's $(1,D,Z,DZ)$. This is in contrast
to Y-CRF1 estimable with OLS of $Y$ on $(1,D,Z,DZ)$.\bigskip

\textbf{Remark 1. }With the slopes in Y-CRF2 specified as linear functions
of $X$, IVE can be applied to Y-CRF2, where the interaction terms between $X$
and $(D,M,DZ)$ are instrumented by the interaction terms between $X$ and $%
(D,Z,DZ)$. This may, however, entail weak IV problems because a single IV $Z$
generates many IV's.\bigskip

\textbf{Remark 2.} If not for C(d), $\{\beta _{z}(X)-\beta _{m}(X)\alpha
_{z}(X)\}\cdot Z\neq 0$ would appear in Y-CRF2:%
\begin{eqnarray}
&&\{\beta _{z}(X)-\beta _{m}(X)\alpha _{z}(X)\}\cdot
Z=[E\{(Y^{01}-Y^{00})(M^{01}-M^{00})|X\}  \notag \\
&&\ \ \ \ \ \ \ \ \ \ \ \ -E(Y^{01}-Y^{00}|M^{10}-M^{00}=1,X)\cdot
E(M^{01}-M^{00}|X)]\cdot Z  \notag \\
&=&\{E(Y^{01}-Y^{00}|M^{01}-M^{00}=1,X)-E(Y^{01}-Y^{00}|M^{10}-M^{00}=1,X)\}
\notag \\
&&\ \ \ \ \cdot E(M^{01}-M^{00}|X)\cdot Z  \TCItag{2.6}
\end{eqnarray}%
as the appendix proof of Theorem 3 reveals. Then the IVE would fail, because
there would be five regressors $(1,D,M,Z,DZ)$, but only four IV's $%
(1,D,Z,DZ) $. C(d) makes the IVE work by removing $Z$ from Y-CRF2, in which
sense C(d) is \textquotedblleft fundamental\textquotedblright .\bigskip

\textbf{Remark 3} (Remark 2 continued). Intuitively speaking, the desired
versus identified indirect effects are (2.3). The $M$ change is induced by $%
D $ in the first expression of (2.3), which is not easy to find due to the $%
M $ endogeneity. In contrast, the $M$ change is induced by $Z$ in the second
expression of (2.3), which is an exogenous change, but not exactly what is
desired. The assumption C(d) is that the latter can be taken as the former,
so that (2.6) is zero and our (or any) IV-based approach works.\bigskip

\textbf{Remark 4. }The slope of $D$ in Y-CRF2 is $Z$-dependent, which
becomes $\beta _{d}(X)-\beta _{m}(X)\alpha _{d}(X)$ when $Z=0$. This is `the
total effect $\beta _{d}(X)$ with $z=0$' minus `the indirect effect with $%
z=0 $' in (2.5), which is thus `the direct effect $E(Y^{10}-Y^{00}+\Delta
Y^{\pm }M^{10}|X)$ with $z=0$' in (1.4) broadly including the interaction
effect $\Delta Y^{\pm }M^{10}$; `$|X$' is not explicit in (1.4).
Analogously, when $Z=1$, the appendix proves under C(e) that the slope of $D$
in Y-CRF2 is the direct effect $E(Y^{10}-Y^{00}+\Delta Y^{\pm }M^{11}|X)$
with $z=1$ in (1.4).

\section{Effect Estimators}

\qquad This section presents effect estimators based on linear
approximations to the unknown functions of $X$ in Y-CRF1 and Y-CRF2. The
total effect is then found with OLS to Y-CRF1, and the direct effect with
IVE to Y-CRF2; the difference between the two effects is the indirect
effect. Linear approximations are restrictive, but they are applied to the
RF functions in the CRF's, not to SF's as in many other empirical studies,
and thus the misspecification issue is less worrisome. Functions of $X$ can
be also used in linear approximations, but for simplicity, we use the same
notation $X$.

\qquad OLS to Y-CRF1 is straightforward to implement, but IVE\ to Y-CRF2 is
not, because $XM$ has to be instrumented by $XZ$. E.g., if $X$ is
10-dimensional, then 10 variables in $XM$ are instrumented by 10 IV's in $XZ$%
: only a single binary IV $Z$ generates 10 IV's, which can be problematic.
Hence, we explore estimators alleviating this dimension problem in the
second half of this section, which use the \textquotedblleft instrument
score\textquotedblright .

\subsection{Estimators with Linear Approximations in X}

\qquad Let $X$ be of dimension $k\times 1$. Linearly approximate all $\beta
(X)$ in Y-CRF1:%
\begin{eqnarray}
&&\ Y=\beta _{0}^{\prime }X+\beta _{d}^{\prime }XD+\beta _{z}^{\prime
}XZ+\beta _{dz}^{\prime }XDZ+U_{1}=\beta _{1}^{\prime }Q_{1}+U_{1}, 
\TCItag{3.1} \\
&&\ \beta _{1}\equiv (\beta _{0}^{\prime },\beta _{d}^{\prime },\beta
_{z}^{\prime },\beta _{dz}^{\prime })^{\prime },\ \ \ \ \ Q_{1}\equiv
(X^{\prime },X^{\prime }D,X^{\prime }Z,X^{\prime }DZ)^{\prime };  \notag
\end{eqnarray}%
e.g., $\beta _{0}^{\prime }X$ is for $\beta _{0}(X)\equiv
E\{Y^{00}+(Y^{01}-Y^{00})M^{00}|X\}$. Do analogously for Y-CRF2:%
\begin{eqnarray}
&&\ Y=\gamma _{0}^{\prime }X+\gamma _{d}^{\prime }XD+\gamma _{m}^{\prime
}XM+\gamma _{dz}^{\prime }XDZ+U_{2}=\gamma _{2}^{\prime }Q_{2}+U_{2}, 
\TCItag{3.2} \\
&&\ \gamma _{2}\equiv (\gamma _{0}^{\prime },\gamma _{d}^{\prime },\gamma
_{m}^{\prime },\gamma _{dz}^{\prime })^{\prime },\ \ \ \ \ Q_{2}\equiv
(X^{\prime },X^{\prime }D,X^{\prime }M,X^{\prime }DZ)^{\prime };  \notag
\end{eqnarray}%
e.g., $\gamma _{d}^{\prime }X$ is for the slope $\beta _{d}(X)-\beta
_{m}(X)\alpha _{d}(X)$ of $D$ with $Z=0$ in Y-CRF2.

\qquad We present the effect estimators based on (3.1) and (3.2) in Theorem
4 below, where we condition on $\bar{X}$ and $\bar{Z}$ as in Lee (2024); an
upper bar denotes the sample average. This is to ignore the errors $%
\overline{X}-E(X)$ and $\bar{Z}-E(Z)$. What is gained by conditioning on $%
\bar{X}$ and $\bar{Z}$ is simplicity in asymptotic inference, and what is
lost is some \textquotedblleft external validity\textquotedblright , as the
findings conditioned on $\bar{X}$ and $\bar{Z}$ apply only to $(X,Z)$-fixed
designs in principle. However, as the simulation study later demonstrates,
not accounting for those errors makes hardly any difference. Let $0_{a\times
b}$ be the $a\times b$ null vector; $\hat{\beta}_{1}$ denotes OLS to Y-CRF1,
and $\hat{\gamma}_{2}$ denotes IVE to Y-CRF2 with $XM$ instrumented by $XZ$.
The proof of Theorem 4 next is omitted, as it is based on linear
combinations of OLS and IVE.\bigskip

\textbf{THEOREM 4.} \textit{(i) The total effect estimator from OLS }$\hat{%
\beta}_{1}\equiv (\hat{\beta}_{0}^{\prime },\hat{\beta}_{d}^{\prime },\hat{%
\beta}_{z}^{\prime },\hat{\beta}_{dz}^{\prime })^{\prime }$ \textit{to
Y-CRF1\ in (3.1) is the linear combination }$\bar{X}^{\prime }\hat{\beta}%
_{d}+\overline{X^{\prime }Z}\hat{\beta}_{dz}$\textit{\ of }$\hat{\beta}_{1}$%
\textit{, from which we have:}%
\begin{eqnarray*}
&&\sqrt{N}\{\bar{X}^{\prime }(\hat{\beta}_{d}-\beta _{d})+\overline{%
X^{\prime }Z}(\hat{\beta}_{dz}-\beta _{dz})\}\rightarrow ^{d}N(0,\Lambda
_{1}),\ \ \ \hat{\Lambda}_{1}\equiv \frac{1}{N}\sum_{i}\hat{\lambda}%
_{1i}^{2}\rightarrow ^{p}\Lambda _{1}\text{ \ where} \\
&&\hat{\lambda}_{1i}\equiv \hat{G}(\frac{1}{N}\sum_{i}Q_{1i}Q_{1i}^{\prime
})^{-1}Q_{1i}\hat{U}_{1i},\ \ \ \hat{G}\equiv (0_{1\times k},\ \bar{X}%
^{\prime },\ 0_{1\times k},\ \overline{X^{\prime }Z}),\ \ \ \hat{U}%
_{1i}\equiv Y_{i}-\hat{\beta}_{1}^{\prime }Q_{1i}.
\end{eqnarray*}%
\textit{(ii) The direct effect estimator from IVE }$\hat{\gamma}_{2}\equiv (%
\hat{\gamma}_{0}^{\prime },\hat{\gamma}_{d}^{\prime },\hat{\gamma}%
_{m}^{\prime },\hat{\gamma}_{dz}^{\prime })^{\prime }$ \textit{to Y-CRF2 in
(3.2) is the linear combination }$\bar{X}^{\prime }\hat{\gamma}_{d}+%
\overline{X^{\prime }Z}\hat{\gamma}_{dz}$\textit{\ of }$\hat{\gamma}_{2}$%
\textit{, from which we have:}%
\begin{eqnarray*}
&&\sqrt{N}\{\bar{X}^{\prime }(\hat{\gamma}_{d}-\gamma _{d})+\overline{%
X^{\prime }Z}(\hat{\gamma}_{dz}-\gamma _{dz})\}\rightarrow ^{d}N(0,\Lambda
_{2}),\ \ \ \ \ \hat{\Lambda}_{2}\equiv \frac{1}{N}\sum_{i}\hat{\lambda}%
_{2i}^{2}\rightarrow ^{p}\Lambda _{2}, \\
&&\ \text{where \ \ }\hat{\lambda}_{2i}\equiv \hat{G}(\frac{1}{N}%
\sum_{i}Q_{1i}Q_{2i}^{\prime })^{-1}Q_{1i}\hat{U}_{2i},\ \ \ \hat{U}%
_{2i}\equiv Y_{i}-\hat{\gamma}_{2}^{\prime }Q_{2i}.
\end{eqnarray*}

\textit{(iii) The indirect effect estimator is }$\bar{X}^{\prime }\hat{\beta}%
_{d}+\overline{X^{\prime }Z}\hat{\beta}_{dz}-(\bar{X}^{\prime }\hat{\gamma}%
_{d}+\overline{X^{\prime }Z}\hat{\gamma}_{dz})$\textit{, and}%
\begin{equation*}
\sqrt{N}\{\ \bar{X}^{\prime }(\hat{\beta}_{d}-\beta _{d})+\overline{%
X^{\prime }Z}(\hat{\beta}_{dz}-\beta _{dz})-\bar{X}^{\prime }(\hat{\gamma}%
_{d}-\gamma _{d})-\overline{X^{\prime }Z}(\hat{\gamma}_{dz}-\gamma _{dz})\ \}
\end{equation*}%
\textit{is asymptotically normal with its variance estimable by }$%
N^{-1}\sum_{i}(\hat{\lambda}_{1i}-\hat{\lambda}_{2i})^{2}$\textit{.\bigskip }

\qquad There are two concerns in the above estimators. One is the
multicollinearity problem due to the same $X$ appearing in all four parts of 
$Q_{1}$ and $Q_{2}$; i.e., all four parts can be highly collinear. The other
concern is weak IV's due to the single binary IV $Z$ generating the multiple
IV's $XZ$ for the possibly endogenous vector $XM$.

\subsection{Estimators for Randomized D Using Instrument Score}

\qquad To overcome the two concerns noted just above, define three
\textquotedblleft scores\textquotedblright :%
\begin{equation}
\mu _{X}\equiv (\pi _{X},\zeta _{X},\xi _{X})^{\prime },\text{ \ \ \ }\pi
_{X}\equiv E(D|X),\text{ \ }\zeta _{X}\equiv E(Z|X),\ \ \xi _{X}\equiv
E(DZ|X);  \tag{3.3}
\end{equation}%
$\pi _{X}$ is the propensity score and $\zeta _{X}$ is the instrument score
(IS). Since $D=0,1$ and $Z=0,1$ generate four cells, $P(D=d,Z=z|X)$ for $%
d,z=0,1$ is equivalent to $\mu _{X}$.

\qquad Letting $1[A]\equiv 1$ if $A$ holds and $0$ otherwise, we can
generalize the dimension reduction idea of Rosenbaum and Rubin (1983) for $D$
to $(D,Z)$: due to C(a),%
\begin{eqnarray*}
&&P(D=d,Z=z|Y^{dm},M^{dz},\ d,m,z=0,1,\mu _{X}) \\
&=&E\{\ E(1[D=d,Z=z]|Y^{dm},M^{dz},\ d,m,z=0,1,X)\ |Y^{dm},M^{dz},\
d,m,z=0,1,\mu _{X}\ \} \\
&=&E\{E(1[D=d,Z=z]|X)\ |Y^{dm},M^{dz},\ d,m,z=0,1,\mu _{X}\}=P(D=d,Z=z\ |\mu
_{X}).
\end{eqnarray*}%
The first and the last expressions establish the key point:%
\begin{equation}
(D,Z)\amalg (Y^{dm},M^{dz},\ d,m,z=0,1)|X\Longrightarrow (D,Z)\amalg
(Y^{dm},M^{dz},\ d,m,z=0,1)|\mu _{X}.  \tag{3.4}
\end{equation}

\qquad Using this, we obtain Y-CRF1 and Y-CRF2 with $X$ replaced with $\mu
_{X}$, and their unknown functions of $\mu _{X}$ can be approximated by
power functions of $\mu _{X}$. Although this alleviates the $X$-dimension
problem, it does not quite solve it because using the first-order terms of $%
\mu _{X}$ entails three IV's $(\pi _{X}Z,\zeta _{X}Z,\xi _{X}Z)$, and using
the second-order terms of $\mu _{X}$ entails as many as nine IV's:%
\begin{equation*}
\pi _{X}Z,\ \pi _{X}^{2}Z,\ \ \zeta _{X}Z,\ \zeta _{X}^{2}Z,\ \ \xi _{X}Z,\
\xi _{X}^{2}X,\ \ \pi _{X}\zeta _{X}Z,\ \pi _{X}\xi _{X}Z,\ \zeta _{X}\xi
_{X}Z.
\end{equation*}%
When $D$ is randomized, however, conditioning on $\zeta _{X}$ is enough,
which we do henceforth.

\qquad With a randomized $D$ and error terms $U_{3}$ and $U_{4}$ satisfying $%
E(U_{3}|D,Z,\zeta _{X})=E(U_{4}|D,Z,\zeta _{X})=0$, we have new Y-CRF1 and
Y-CRF2, instead of the original CRF's:%
\begin{eqnarray}
&&\ Y=\beta _{0}(\zeta _{X})+\beta _{d}(\zeta _{X})D+\beta _{z}(\zeta
_{X})Z+\beta _{dz}(\zeta _{X})DZ\ +U_{3},  \TCItag{3.5} \\
&&\ Y=\beta _{0}(\zeta _{X})-\beta _{m}(\zeta _{X})\alpha _{0}(\zeta
_{X})+\{\beta _{d}(\zeta _{X})-\beta _{m}(\zeta _{X})\alpha _{d}(\zeta
_{X})\}D  \notag \\
&&\ \ \ \ \ \ +\beta _{m}(\zeta _{X})M+\{\beta _{dz}(\zeta _{X})-\beta
_{m}(\zeta _{X})\alpha _{dz}(\zeta _{X})\}DZ\ +U_{4}.  \TCItag{3.6}
\end{eqnarray}%
We now present effect estimators incorporating this dimension reduction
idea.\bigskip

\qquad Let $\zeta _{X}=\Phi (X^{\prime }\theta )$ for a parameter $\theta $,
and $\hat{\zeta}_{X}=\Phi (X^{\prime }\hat{\theta})$ be the probit
regression estimator of $Z$ on $X$; $\Phi (\cdot )$ is the $N(0,1)$
distribution function. As it is simpler to condition on $X^{\prime }\hat{%
\theta}$ instead of $\hat{\zeta}_{X}$, define:%
\begin{equation*}
W_{\theta }\equiv \{1,\ (X^{\prime }\theta ),\ (X^{\prime }\theta )^{2},\
...,\ (X^{\prime }\theta )^{J}\}^{\prime }.
\end{equation*}%
Then, instead of (3.5) and (3.6), we consider:%
\begin{eqnarray}
&&Y=\beta _{0}^{\prime }W_{\theta }+\beta _{d}^{\prime }W_{\theta }D+\beta
_{z}^{\prime }W_{\theta }Z+\beta _{dz}W_{\theta }DZ+U_{3}=\beta _{1}^{\prime
}Q_{1}(\theta )+U_{3},  \TCItag{3.7} \\
&&Y=\gamma _{0}^{\prime }W_{\theta }+\gamma _{d}W_{\theta }D+\gamma
_{m}^{\prime }W_{\theta }M+\gamma _{dz}W_{\theta }DZ+U_{4}=\gamma
_{2}^{\prime }Q_{2}(\theta )+U_{4},  \TCItag{3.8} \\
&&\underset{4(J+1)\times 1}{\beta _{1}}\equiv (\beta _{0}^{\prime },\beta
_{d}^{\prime },\beta _{z}^{\prime },\beta _{dz}^{\prime })^{\prime },\ \ \ 
\underset{4(J+1)\times 1}{Q_{1}(\theta )}\equiv (W_{\theta }^{\prime },\
W_{\theta }^{\prime }D,\ W_{\theta }^{\prime }Z,\ W_{\theta }^{\prime
}DZ)^{\prime },  \notag \\
&&\underset{4(J+1)\times 1}{\gamma _{2}}\equiv (\gamma _{0}^{\prime },\gamma
_{d}^{\prime },\gamma _{m}^{\prime },\gamma _{dz}^{\prime })^{\prime },\ \ \ 
\underset{4(J+1)\times 1}{Q_{2}(\theta )}\equiv (W_{\theta }^{\prime },\
W_{\theta }^{\prime }D,\ W_{\theta }^{\prime }M,\ W_{\theta }^{\prime
}DZ)^{\prime };  \notag
\end{eqnarray}%
to save notation, (3.7) and (3.8) use the same notation $\beta $'s and $%
\gamma $'s as in (3.1) and (3.2), and the numbers below a matrix denotes its
dimension.

\qquad For $W_{\theta }$, $J=1$ can allow only a monotonic function of $%
X^{\prime }\theta $, and thus we recommend $J=2$ or $J=3$; going beyond $J=3$
may not be a good idea due to the multicollinearity problem. The proof for
Theorem 5 next is omitted, which conditions on all $X_{i}$'s and $Z_{i}$'s
to fix $\hat{\theta}$, not just on $\bar{X}$ and $\bar{Z}$, differently from
Theorem 4.\bigskip

\textbf{THEOREM 5.} \textit{(i) The total effect estimator from OLS }$\tilde{%
\beta}_{1}\equiv (\tilde{\beta}_{0}^{\prime },\tilde{\beta}_{d}^{\prime },%
\tilde{\beta}_{z}^{\prime },\tilde{\beta}_{dz}^{\prime })^{\prime }$\textit{%
\ to (3.7) is the linear combination }$\overline{W_{\hat{\theta}}}^{\prime }%
\tilde{\beta}_{d}+\overline{W_{\hat{\theta}}Z}^{\prime }\tilde{\beta}_{dz}$%
\textit{\ of }$\tilde{\beta}_{1}$, \textit{from which we have:}%
\begin{eqnarray*}
&&\ \sqrt{N}\{\overline{W_{\hat{\theta}}}^{\prime }(\tilde{\beta}_{d}-\beta
_{d})+\overline{W_{\hat{\theta}}Z}^{\prime }(\tilde{\beta}_{dz}-\beta
_{dz})\}\rightarrow ^{d}N(0,\Omega _{1}),\ \ \ \tilde{\Omega}_{1}\equiv 
\frac{1}{N}\sum_{i}\tilde{\lambda}_{1i}^{2}\rightarrow ^{p}\Omega _{1}, \\
&&\ \tilde{\lambda}_{1i}\equiv \tilde{G}\{\frac{1}{N}\sum_{i}Q_{1i}(\hat{%
\theta})Q_{1i}^{\prime }(\hat{\theta})\}^{-1}Q_{1i}(\hat{\theta})\tilde{U}%
_{3i},\ \ \ \tilde{G}\equiv (0_{1\times (J+1)},\ \overline{W_{\hat{\theta}}}%
^{\prime },\ 0_{1\times (J+1)},\ \overline{W_{\hat{\theta}}Z}^{\prime }), \\
&&\ \tilde{U}_{3i}\equiv Y_{i}-\tilde{\beta}_{1}^{\prime }Q_{1i}(\hat{\theta}%
).
\end{eqnarray*}%
\textit{(ii) The direct effect estimator from IVE }$\tilde{\gamma}_{2}\equiv
(\tilde{\gamma}_{0}^{\prime },\tilde{\gamma}_{d}^{\prime },\tilde{\gamma}%
_{m}^{\prime },\tilde{\gamma}_{dz}^{\prime })^{\prime }$\textit{\ to (3.8)
is the linear combination} $\overline{W_{\hat{\theta}}}^{\prime }\tilde{%
\gamma}_{d}+\overline{W_{\hat{\theta}}Z}^{\prime }\tilde{\gamma}_{dz}$%
\textit{\ of }$\tilde{\gamma}_{2}$\textit{, from which we have:}%
\begin{eqnarray*}
&&\sqrt{N}\{\overline{W_{\hat{\theta}}}^{\prime }(\tilde{\gamma}_{d}-\gamma
_{d})+\overline{W_{\hat{\theta}}Z}^{\prime }(\tilde{\gamma}_{dz}-\gamma
_{dz})\}\rightarrow ^{d}N(0,\Omega _{2}),\ \ \ \ \ \tilde{\Omega}_{2}\equiv 
\frac{1}{N}\sum_{i}\tilde{\lambda}_{2i}^{2}\rightarrow ^{p}\Omega _{2}, \\
&&\ \text{where \ \ \ \ }\tilde{\lambda}_{2i}\equiv \tilde{G}\frac{1}{N}%
\sum_{i}Q_{1i}(\hat{\theta})Q_{2i}^{\prime }(\hat{\theta})\}^{-1}Q_{1i}(\hat{%
\theta})\tilde{U}_{4i},\ \ \ \ \ \tilde{U}_{4i}\equiv Y_{i}-\tilde{\gamma}%
_{2}^{\prime }Q_{2i}(\hat{\theta}).
\end{eqnarray*}

\textit{(iii) The indirect effect estimator is }$\overline{W_{\hat{\theta}}}%
^{\prime }\tilde{\beta}_{d}+\overline{W_{\hat{\theta}}Z}^{\prime }\tilde{%
\beta}_{dz}-\overline{W_{\hat{\theta}}}^{\prime }\tilde{\gamma}_{d}-%
\overline{W_{\hat{\theta}}Z}^{\prime }\tilde{\gamma}_{dz}$\textit{, and }%
\begin{equation*}
\sqrt{N}\{\ \overline{W_{\hat{\theta}}}^{\prime }(\tilde{\beta}_{d}-\beta
_{d})+\overline{W_{\hat{\theta}}Z}^{\prime }(\tilde{\beta}_{dz}-\beta _{dz})-%
\overline{W_{\hat{\theta}}}^{\prime }(\tilde{\gamma}_{d}-\gamma _{d})-%
\overline{W_{\hat{\theta}}Z}^{\prime }(\tilde{\gamma}_{dz}-\gamma _{dz})\ \}
\end{equation*}%
\textit{is asymptotically normal with its variance estimable by }$%
N^{-1}\sum_{i}(\tilde{\lambda}_{1i}-\tilde{\lambda}_{2i})^{2}$\textit{.}

\section{Simulation Study}

\qquad Our base design is the following, where $D$ is randomized with $%
P(D=0)=P(D=1)=0.5$, $N=1000$ or$\ 4000$, and $10000$ simulation repetitions:%
\begin{eqnarray*}
&&\ Z=1[0<\vartheta _{1}+\vartheta _{x}X_{0}+e],\ \ \ X_{0}\sim N(0,1),\ \ \
e\sim N(0,1)\amalg X_{0},\ \ \ \vartheta _{1}=0,\ \ \vartheta _{x}=1; \\
&&M^{dz}=1[0.5<\alpha _{1}+\alpha _{d}d+\alpha _{z}z+\alpha
_{x}X_{0}+\varepsilon ],\ \varepsilon \sim N(0,1)\amalg (X_{0},e), \\
&&\ \ \ \ \ \ \ \ \ \ \alpha _{1}=0,\ \ \ \alpha _{d}=1,\ \ \ \alpha
_{z}=1,\ \ \ \alpha _{x}=1;\text{ }Y\text{ is continuous or binary with} \\
&&Y^{dm\ast }=\beta _{0}+\beta _{d}d+\beta _{m}m+\beta _{dm}dm+\beta
_{x}X_{0}+U,\ \ \ Y^{dm}=Y^{dm\ast }\text{ or }1[0.5<Y^{dm\ast }], \\
&&\ \ \ \ \ \ \ \ \ \ \beta _{0}=0,\ \ \beta _{d}=0.5,\ \ \beta _{m}=1,\ \
\beta _{dm}=0.5,\ \ \beta _{x}=1, \\
&&U\sim N(0,1)\amalg (X_{0},e,\varepsilon )\text{ for exogenous }M\text{, \ }%
U=N(0,1)+\varepsilon \text{ for endogenous }M;
\end{eqnarray*}%
$U=N(0,1)+\varepsilon $ is standardized, where $SD$ stands for standard
deviation

\qquad Then we generate $M$ with (2.1), and $Y$ with%
\begin{equation*}
Y=(1-D)(1-M)Y^{00}+(1-D)MY^{01}+D(1-M)Y^{10}+DMY^{11}.
\end{equation*}%
The total effect is calculated as the sample-mean version of (1.4) at each
run, and the direct effect as the sample-mean version of M2M excluding $%
(Y^{01}-Y^{00})(M^{10}-M^{00})$ and $(Y^{01}-Y^{00})(M^{11}-M^{01})$.

\qquad We try four designs, depending on continuous/binary $Y$ and
exogenous/endogenous $M$. Occasionally, the simulation run stops due to a
singular matrix problem, in which case the run is aborted and the simulation
data are redrawn. Also, as will be seen shortly, sometimes outliers occur
which distort the entire simulation results when $N=1000$, but this problem
disappears when $N=4000$.

\begin{center}
\begin{tabular}{ccccc}
\hline\hline
\multicolumn{5}{c}{Table 1. Continuous $Y$: \TEXTsymbol{\vert}BIAS/effect%
\TEXTsymbol{\vert}, simSD/\TEXTsymbol{\vert}effect\TEXTsymbol{\vert} (RMSE/%
\TEXTsymbol{\vert}effect\TEXTsymbol{\vert}), AsySD/\TEXTsymbol{\vert}effect%
\TEXTsymbol{\vert}} \\ 
& Exo M, N=1000 & Exo M, N=4000 & Endo M, N=1000 & Endo M, N=4000 \\ \hline
\multicolumn{5}{c}{OLS for exogenous $M$} \\ 
\multicolumn{1}{l}{tot} & \multicolumn{1}{l}{\small .017 .066 (.069) .066} & 
\multicolumn{1}{l}{\small .00 .033 (.033) .033} & \multicolumn{1}{l}{\small %
.01 .075 (0.076) .075} & \multicolumn{1}{l}{\small .01 .038 (.038) .038} \\ 
\multicolumn{1}{l}{dir} & \multicolumn{1}{l}{\small .00 .080 \ (.080) .080}
& \multicolumn{1}{l}{\small .00 .033 (.033) .033} & \multicolumn{1}{l}%
{\small .25 .073 (0.27) .073} & \multicolumn{1}{l}{\small .26 .037 (.26) .036%
} \\ 
\multicolumn{1}{l}{ind} & \multicolumn{1}{l}{\small .078 .18 \ (.19) \ .18}
& \multicolumn{1}{l}{\small .045 .085 (.096) .085} & \multicolumn{1}{l}%
{\small 1.1 \ .26 \ (1.1) \ \ .26} & \multicolumn{1}{l}{\small 1.1 \ \ .13
(1.1) \ \ .13} \\ 
\multicolumn{5}{c}{IVE$_{1}$ for endogenous $M$ controlling $X$} \\ 
\multicolumn{1}{l}{tot} & \multicolumn{1}{l}{\small .017 .066 (.068) .066} & 
\multicolumn{1}{l}{\small .00 .032 (.032) .032} & \multicolumn{1}{l}{\small %
.01 .075 (0.075) .074} & \multicolumn{1}{l}{\small .01\ .037 (.038) .037} \\ 
\multicolumn{1}{l}{dir} & \multicolumn{1}{l}{\small .00 \ \ \ .11 (.11) \ \
.11} & \multicolumn{1}{l}{\small .010 .054 (.055) .054} & \multicolumn{1}{l}%
{\small .01 \ \ .12 (0.12) \ .11} & \multicolumn{1}{l}{\small .00 .055
(.055) .055} \\ 
\multicolumn{1}{l}{ind} & \multicolumn{1}{l}{\small .086 \ \ .38 (.39) \ \
.37} & \multicolumn{1}{l}{\small .045\ .18 \ (.18) \ \ .18} & 
\multicolumn{1}{l}{\small .01 \ \ .35 (0.35) \ .36} & \multicolumn{1}{l}%
{\small .028\ .17 (.17) \ .17} \\ 
\multicolumn{5}{c}{IVE$_{2}$ for endogenous $M$ controlling $(\zeta
_{X},\zeta _{X}^{2})$} \\ 
tot & \multicolumn{1}{l}{\small .017 .066 (.068) .065} & \multicolumn{1}{l}%
{\small .00 .032 (.032) .032} & \multicolumn{1}{l}{\small .01 .075 (0.075)
.074} & \multicolumn{1}{l}{\small .01 .037 (.038) .037} \\ 
dir & \multicolumn{1}{l}{\small .00 \ \ \ .13 (.13) \ \ .13} & 
\multicolumn{1}{l}{\small .010 .061 (.062) .061} & \multicolumn{1}{l}{\small %
.021 .14 (0.14) \ .14} & \multicolumn{1}{l}{\small .00 .062 (.062) .062} \\ 
ind & \multicolumn{1}{l}{\small .089 \ .48 (.49) \ .48} & \multicolumn{1}{l}%
{\small .045 \ .21 (.22) \ \ .21} & \multicolumn{1}{l}{\small .051\ .47
(0.48) \ .46} & \multicolumn{1}{l}{\small .017 \ .21 (.21) \ .21} \\ 
\multicolumn{5}{c}{IVE$_{3}$ for endogenous $M$ controlling $(\zeta
_{X},\zeta _{X}^{2},\zeta _{X}^{3})$} \\ 
tot & \multicolumn{1}{l}{\small .017 .066 (.068) .065} & \multicolumn{1}{l}%
{\small .00 .032 (.032) .032} & \multicolumn{1}{l}{\small .01 .075 (0.075)
.074} & \multicolumn{1}{l}{\small .01 .037 (.038) .037} \\ 
dir & \multicolumn{1}{l}{\small .01 \ \ \ .79 (.79) \ \ 2.0} & 
\multicolumn{1}{l}{\small .010 .068 (.069) .068} & \multicolumn{1}{l}{\small %
.040 2.1 (2.1) 20} & \multicolumn{1}{l}{\small .01 .070 (.070) .070} \\ 
ind & \multicolumn{1}{l}{\small .12 \ \ \ 3.4 (3.4) \ 8.6} & 
\multicolumn{1}{l}{\small .045 \ .25 (.25) \ .25} & \multicolumn{1}{l}%
{\small .13 \ \ 8.5 (8.5) \ \ 82} & \multicolumn{1}{l}{\small .00 \ \ .24
(.24) \ \ .25} \\ \hline
\multicolumn{5}{c}{tot: total effect; \ dir: direct; ind: indirect; 0 to the
left of decimal point omitted;} \\ 
\multicolumn{5}{c}{2 significant figures mostly, except for rounded numbers 
\TEXTsymbol{<} 0.00; simSD is simulation} \\ 
\multicolumn{5}{c}{SD; AsySD is the average of the asymptotic SD's based on
Theorem 4 or\ 5} \\ \hline\hline
\end{tabular}
\end{center}

\qquad Table 1 presents the results for continuous $Y$ with exogenous $M$ on
the left-hand side and endogenous $M$ on the right-hand side. Each entry has
four numbers: \TEXTsymbol{\vert}BIAS\TEXTsymbol{\vert}, simulation (i.e.,
the true) SD (\textquotedblleft simSD\textquotedblright ), root mean squared
error (RMSE), and the average of $10000$ asymptotic SD's (\textquotedblleft
asySD\textquotedblright ) to see how accurate the variance formulas in
Theorems 4 and 5 are, compared with the true simulation SD. Since the
effects vary across the designs, we divide each number by the absolute
effect magnitude for standardization. IVE$_{1}$ is the IVE controlling $X$,
not $\zeta _{X}$; IVE$_{2}$ is the IVE controlling $(\zeta _{X},\zeta
_{X}^{2})$; and IVE$_{3}$ is the IVE\ controlling $(\zeta _{X},\zeta
_{X}^{2},\zeta _{X}^{3})$. No dimension problem occurs in our designs
because there is only one regressor $X_{0}$, but it is still of interest to
see how controlling $\zeta _{X}$ works relative to controlling $X $.

\qquad The left half of Table 1 with exogenous $M$ shows the performance
ranking: with `$\succ $' standing for \textquotedblleft better than in terms
of RMSE\textquotedblright ,%
\begin{equation}
\text{OLS}\succ \text{IVE}_{1}\succ \text{IVE}_{2}\succ \text{IVE}_{3}. 
\tag{4.1}
\end{equation}%
The aforementioned outlier problem can be seen in IVE$_{3}$ with $N=1000$,
as its SD 3.4 is almost 10 times higher than the SD's of the other
estimators. However, the problem disappears with $N=4000$. The right half of
Table 1 with endogenous $M$ shows that OLS is highly biased, which persists
even when $N=4000$, whereas all three IVE's perform well with near-zero
biases. The ranking among the IVE's are the same as in (4.1). Except for IVE$%
_{3}$ with $N=1000$ in Table 1, the asymptotic SD's are almost the same as
the corresponding simulation SD's to show that Theorems 4 and 5 work well.

\qquad The structure of Table 2 is the same as that of Table 1, except for $%
Y $ being binary. The left half of Table 2 with exogenous $M$ shows that the
performance ranking with $N=4000$ is roughly that%
\begin{equation}
\text{OLS}\succ \text{IVE}_{1}\simeq \text{IVE}_{2}\simeq \text{IVE}_{3} 
\tag{4.2}
\end{equation}%
although IVE$_{3}$ performs clearly worse than IVE$_{1}$ and IVE$_{2}$ with $%
N=1000$. The right half of Table 1 with endogenous $M$ shows that OLS is
highly biased, which persists even when $N=4000$, whereas all three IVE's
perform relatively better. The performance ranking is almost the reverse of
(4.1):%
\begin{equation}
\text{IVE}_{2}\succ \text{IVE}_{3}\succ \text{IVE}_{1}\succ \text{OLS} 
\tag{4.3}
\end{equation}%
although IVE$_{3}$ performs noticeably poorly due to outliers when $N=1000$.
When $N=4000$, IVE$_{2}$ and IVE$_{3}$ perform clearly better than IVE$_{1}$
despite no dimension problem in $X$.

\begin{center}
\begin{tabular}{ccccc}
\hline\hline
\multicolumn{5}{c}{Table 2. Binary $Y$: \TEXTsymbol{\vert}BIAS/effect%
\TEXTsymbol{\vert}, simSD/\TEXTsymbol{\vert}effect\TEXTsymbol{\vert} (RMSE/%
\TEXTsymbol{\vert}effect\TEXTsymbol{\vert}), AsySD/\TEXTsymbol{\vert}effect%
\TEXTsymbol{\vert}} \\ 
& Exo M, N=1000 & Exo M, N=4000 & Endo M, N=1000 & Endo M, N=4000 \\ \hline
\multicolumn{5}{c}{OLS for exogenous $M$} \\ 
\multicolumn{1}{l}{tot} & \multicolumn{1}{l}{\small .01 .10 (.10) \ .10} & 
\multicolumn{1}{l}{\small .019 .052 (.055) .052} & \multicolumn{1}{l}{\small %
.070 .11 (.13) .10} & \multicolumn{1}{l}{\small .021 .055 (.059) .055} \\ 
\multicolumn{1}{l}{dir} & \multicolumn{1}{l}{\small .035 .16 (.16) .16} & 
\multicolumn{1}{l}{\small .029 .081 (.086) .080} & \multicolumn{1}{l}{\small %
.59 .14 \ (.61) \ .13} & \multicolumn{1}{l}{\small .56 \ .072 (.56) .072} \\ 
\multicolumn{1}{l}{ind} & \multicolumn{1}{l}{\small .10 \ .20 (.23) .20} & 
\multicolumn{1}{l}{\small .12 \ \ \ .10 (.16) \ .10} & \multicolumn{1}{l}%
{\small .60 \ .20 (.64) .20} & \multicolumn{1}{l}{\small .64 \ \ .10 (.65)
.10} \\ 
\multicolumn{5}{c}{IVE for endogenous $M$ controlling $X$} \\ 
\multicolumn{1}{l}{tot} & \multicolumn{1}{l}{\small .00 \ .10 (.10) .10} & 
\multicolumn{1}{l}{\small .019 .051 (.055) .051} & \multicolumn{1}{l}{\small %
.070 .10 (.12) .10} & \multicolumn{1}{l}{\small .021 .054 (.058) .053} \\ 
\multicolumn{1}{l}{dir} & \multicolumn{1}{l}{\small .083 .23 (.25) .23} & 
\multicolumn{1}{l}{\small .077 \ .12 (.14) .12} & \multicolumn{1}{l}{\small %
.22 \ .21 (.30) .21} & \multicolumn{1}{l}{\small .16 \ \ .11 (.20) .11} \\ 
\multicolumn{1}{l}{ind} & \multicolumn{1}{l}{\small .20 \ .42 (.46) \ .42} & 
\multicolumn{1}{l}{\small .23 \ \ \ .20 (.30)\ .21} & \multicolumn{1}{l}%
{\small .12 \ .25 (.27) .25} & \multicolumn{1}{l}{\small .15 \ \ .12 (.20)
.12} \\ 
\multicolumn{5}{c}{IVE for endogenous $M$ controlling $(\zeta _{X},\zeta
_{X}^{2})$} \\ 
tot & \multicolumn{1}{l}{\small .01 .10 (.10) .10} & \multicolumn{1}{l}%
{\small .019 .051 (.054) .050} & \multicolumn{1}{l}{\small .071 .10 (.12) .10%
} & \multicolumn{1}{l}{\small .021 .053 (.057) .053} \\ 
dir & \multicolumn{1}{l}{\small .046 .27 (.27) .27} & \multicolumn{1}{l}%
{\small .051 \ .13 (.14) \ .13} & \multicolumn{1}{l}{\small .11 \ .26 (.28)
.25} & \multicolumn{1}{l}{\small .062 \ .13 \ (.14) \ .13} \\ 
ind & \multicolumn{1}{l}{\small .074 .49 (.50) .49} & \multicolumn{1}{l}%
{\small .048 \ .23 (.24) .23} & \multicolumn{1}{l}{\small .016 .29 (.29) .29}
& \multicolumn{1}{l}{\small .028 \ .14 (.14) .14} \\ 
\multicolumn{5}{c}{IVE for endogenous $M$ controlling $(\zeta _{X},\zeta
_{X}^{2},\zeta _{X}^{3})$} \\ 
tot & \multicolumn{1}{l}{\small .01 \ .10 (.10) \ .10} & \multicolumn{1}{l}%
{\small .019 .050 (.054) .050} & \multicolumn{1}{l}{\small .070 .10 (.12)
.099} & \multicolumn{1}{l}{\small .021 .053 (.057) .052} \\ 
dir & \multicolumn{1}{l}{\small .071 .50 (.51) .84} & \multicolumn{1}{l}%
{\small .061 \ \ .13 (.15) \ .13} & \multicolumn{1}{l}{\small .11 \ \ 3.6\
(3.6) \ 16} & \multicolumn{1}{l}{\small .057 \ .14 (.15) .14} \\ 
ind & \multicolumn{1}{l}{\small .13 \ 1.0 (1.0) 1.7} & \multicolumn{1}{l}%
{\small .071 \ \ .25 (.26) \ .25} & \multicolumn{1}{l}{\small .020 \ 4.6
(4.6) \ 20} & \multicolumn{1}{l}{\small .023 \ .15 (.15) .15} \\ \hline
\multicolumn{5}{c}{tot: total effect; \ dir: direct; ind: indirect; 0 to the
left of decimal point omitted;} \\ 
\multicolumn{5}{c}{2 significant figures mostly, except for rounded numbers 
\TEXTsymbol{<} 0.00; simSD is simulation} \\ 
\multicolumn{5}{c}{SD; AsySD is the average of the asymptotic SD's based on
Theorem 4 or\ 5} \\ \hline\hline
\end{tabular}
\end{center}

\qquad Overall, our simulation study confirms that OLS is much biased when $%
M $ is endogenous. Also, IVE$_{2}$ controlling $(\zeta _{X},\zeta _{X}^{2})$
overall performs at least as well as `IVE$_{1}$ controlling $X$' and `IVE$%
_{3}$ controlling $(\zeta _{X},\zeta _{X}^{2},\zeta _{X}^{3})$'.
Surprisingly, this holds despite no dimension problem in $X$ in our
simulation designs.

\bigskip

\section{Small Class Effects on Test Scores}

\qquad Our empirical analysis uses the Project Star data analyzed in depth
by Krueger (1999), and our data was drawn from Stock and Watson (2007); see
\textquotedblleft
https://search.r-project.org/CRAN/refmans/AER/html/STAR.html%
\textquotedblright\ for the details on the original data and the data in
Stock and Watson (2007).

\qquad The outcome variable is the sum of the math and reading SAT scores in
grade 3, which is denoted as $Y_{3}$, because the grade-2 score $Y_{2}$ and
the grade-1 score $Y_{1}$ are used as well in our analysis. $D$ is being in
a small class or not (of 13-17 pupils, relative to the regular class size
22-25) that was randomized at the school level. The randomization was done
either at kindergarten or grade 1, but we use only the pupils who were
randomized at kindergarten, never to change the treatment status up to grade
3.

\qquad The covariates are: black or not (\textquotedblleft
blk\textquotedblright ), boy or not (\textquotedblleft boy\textquotedblright
), the sum of teaching experiences of the teachers in years
(\textquotedblleft expi\textquotedblright ), and eligibility for free lunch
or not (\textquotedblleft lunch\textquotedblright ) representing the family
income level. Lunch and expi vary across grades, but since our outcome
variable is for grade 3, we use only grade-3 observations for lunch and
expi. In the actual estimation, we transform expi into $\ln ($expi+1$)$, and
use $Y_{3}/SD(Y_{3})$ as the outcome $Y$ to see the effects relative to $%
SD(Y_{3})$. Our working sample size is $N=1991$, and the data are for the
academic years 1985-89 in the state of Tennessee, the U.S.A.

\qquad We set $M=1[Y_{2}$ p-quintile$\ <Y_{2}]$ for the five quintile values
of $p=0.1,\ 0.3,\ 0.5,\ 0.7$ and $0.9$, because $D$ may influence $Y_{3}$
directly as well as indirectly through $Y_{2}$. Since $D$ is randomized, the
endogeneity issue can arise only for $M$. As for the IV $Z$ for $M$, we set $%
Z=1[Y_{1}$ p-quintile\ $<Y_{1}]$, adopting the old saying \textquotedblleft
a boxer is only as good as his last bout\textquotedblright . That is, if the
past scores can affect the current score, only the immediate past score
matters. This means that the IV exclusion restriction holds for $Y_{1}$. The
IV inclusion restriction is also satisfied, because $Y_{1}$ precedes $Y_{2}$%
, and $Cor(Y_{1},Y_{2})=0.77$ whereas $Cor(M,Z)=0.47$ for $p=0.1$ e.g.; $Cor$
stands for correlation.

\qquad Transforming $(Y_{2},Y_{1})$ to binary $(M,Z)$ entails some loss of
information, as the decline in the correlations just above demonstrates.
Nevertheless, the choice of the test score p-quintile values provides a
chance to see how pupils at the different quintiles are affected differently
in their indirect effect through $M$ (i.e., through enhanced $Y_{2}$). A
positive indirect effect can happen, if a higher $Y_{2}$ raises one's
self-esteem and confidence, leading to a higher motivation to study harder
and possibly attracting better peers. Our empirical findings provided
shortly below indeed confirm this conjecture.

\qquad Table 3 shows descriptive statistics for the variables, where average
(SD), minimum and maximum are provided; for dummies, the minimum and maximum
are omitted.

\begin{center}
\begin{tabular}{llll}
\hline\hline
\multicolumn{4}{c}{Table 3. Descriptive Statistics: Average (SD), Min, Max; $%
N=1991$} \\ \hline
$D\ $(small class or not) & 0.33 (0.47) & black & 0.23 (0.42) \\ 
grade-3 test score $Y_{3}$ & 1255 (70), 1044, 1527 & boy & 0.49 (0.50) \\ 
grade-2 test score $Y_{2}$ & 1195 (79), 985, 1431 & free lunch & 0.35 (0.48)
\\ 
grade-1 test score $Y_{1}$ & 1088 (84), 883, 1327 & teacher expi & 13.7
(8.5), 0, 38 \\ \hline
\multicolumn{4}{c}{Min \& Max not shown for dummies; 99\% are blacks or
whites; $M$ ($Z$) is a binary} \\ 
\multicolumn{4}{c}{transform of $Y_{2}$ ($Y_{1}$); teacher expi is the sum
of teachers' experiences in years} \\ \hline\hline
\end{tabular}
\end{center}

\qquad Table 4 presents the effect estimation results, where OLS means the
OLS-based effect decomposition (Lee 2024) for exogenous $M$, and IVE$_{1}$,
IVE$_{2}$ and IVE$_{3}$ are the (OLS- and ) IVE-based effect decompositions
of this paper for endogenous $M$ controlling $X$, $(\zeta _{X},\zeta
_{X}^{2})$ and $(\zeta _{X},\zeta _{X}^{2},\zeta _{X}^{3})$, respectively.
Although the total effects under exogenous $M$ in the OLS column are the
same $0.19$ for all quintiles, their decomposition varies across the
quintiles, with the direct effects ranging over $0.10$ to $0.15$ (i.e.,
these numbers times $SD(Y_{3})$), whereas the indirect effects range over $%
0.042$ to $0.089$, being always smaller than the direct effects. The total
effect $0.19$ in the OLS column is also the same as the simple group mean
difference for $E\{Y_{3}/SD(Y_{3})|D=1\}-E\{Y_{3}/SD(Y_{3})|D=0\}$.

\qquad In Table 4, when endogenous $M$ is allowed for, the total effects
range over $0.062$ to $0.15$, being much smaller than the total effect $0.19$
under exogenous $M$. In the decomposition of the total effect with
endogenous $M$, the indirect effects are not always smaller than the direct
effects; e.g. the indirect effect is greater than the direct effect for the $%
0.3$ and $0.5$ quintiles, although they are not statistically significant.

\qquad In most cases of Table 4, the t-values of IVE$_{1}$ are greater than
those of IVE$_{2}$, which are in turn greater than those of IVE$_{3}$; the
statistical significance of the IVE's at the conventional 5\% level changes
only for 0.7 and 0.9 quintiles at most. The reason for the decreasing
statistical significance is likely to be the multicollinearity among $(\zeta
_{X},\zeta _{X}^{2},\zeta _{X}^{3})$. Other than this, the effects and
t-values are similar across the three IVE's. Note that, since $X=($blk,\
boy,\ expi$_{3}$, free-lunch$_{3})^{\prime }$ is four-dimensional where the
subscript $3$ denotes `grade 3', the dimension reduction is not much: by 2
when $(\zeta _{X},\zeta _{X}^{2})$ are used, and only by 1 when $(\zeta
_{X},\zeta _{X}^{2},\zeta _{X}^{3})$ is used.

\begin{center}
\begin{tabular}{cccccc}
\hline\hline
\multicolumn{6}{c}{Table 4. Effects (tv's) with Outcome $Y_{3}/SD(Y_{3})$:
OLS, IVE$_{1}$, IVE$_{2}$ and IVE$_{3}$} \\ 
Quintile & Effect & OLS & IVE$_{1}$ & IVE$_{2}$ & IVE$_{3}$ \\ \hline
0.1 & \textit{total} & \textit{0.19 (4.18)} & \textit{0.15 (3.41)} & \textit{%
0.15 (3.42)} & \textit{0.15 (3.41)} \\ 
& direct & 0.14 (3.31) & 0.086 (1.88) & 0.087 (1.91) & 0.089 (1.93) \\ 
& indirect & 0.049 (2.90) & 0.059 (1.86) & 0.060 (1.88) & 0.057 (1.73) \\ 
0.3 & \textit{total} & \textit{0.19 (4.19)} & \textit{0.11 (2.82)} & \textit{%
0.10 (2.58)} & \textit{0.10 (2.58)} \\ 
& direct & 0.11 (3.00) & 0.052 (1.23) & 0.045 (1.06) & 0.044 (1.02) \\ 
& indirect & 0.077 (3.05) & 0.060 (1.52) & 0.057 (1.43) & 0.059 (1.47) \\ 
0.5 & \textit{total} & \textit{0.19 (4.18)} & \textit{0.093 (2.43)} & 
\textit{0.082 (2.16)} & \textit{0.081 (2.14)} \\ 
& direct & 0.10 (2.86) & 0.039 (0.98) & 0.030 (0.72) & 0.035 (0.84) \\ 
& indirect & 0.089 (3.08) & 0.054 (1.39) & 0.052 (1.28) & 0.046 (1.15) \\ 
0.7 & \textit{total} & \textit{0.19 (4.18)} & \textit{0.083 (2.14)} & 
\textit{0.067 (1.74)} & \textit{0.062 (1.62)} \\ 
& direct & 0.13 (3.37) & 0.093 (2.23) & 0.078 (1.80) & 0.070 (1.56) \\ 
& indirect & 0.062 (2.32) & -0.010 (-0.28) & -0.011\ (-0.29) & -0.008 (-0.20)
\\ 
0.9 & \textit{total} & \textit{0.19 (4.19)} & \textit{0.13 (2.96)} & \textit{%
0.11 (2.68)} & \textit{0.13 (2.28)} \\ 
& direct & 0.15 (3.64) & 0.089 (1.52) & 0.098 (1.89) & 0.087 (0.62) \\ 
& indirect & 0.042 (2.01) & 0.037 (0.66) & 0.017 (0.37) & 0.042 (0.30) \\ 
\hline
\multicolumn{6}{c}{`p quintile' means $M=1[($p-quintile of $Y_{2})<Y_{2}]$
\& $Z=1[($p-quintile of $Y_{1}<Y_{1})$;} \\ 
\multicolumn{6}{c}{OLS for exo $M$; IVE$_{1}$, IVE$_{2}$ \& IVE$_{3}$
control $X$, $(\zeta _{X},\zeta _{X}^{2})$ \& $\,(\zeta _{X},\zeta
_{X}^{2},\zeta _{X}^{3})$ for endo $M$} \\ \hline\hline
\end{tabular}
\end{center}

\qquad Krueger (1999, p. 514) shows that the effect in the third year is $%
0.19$. This is exactly the same as our finding in Table 2 under exogenous $M$%
, despite that the Krueger's result is based on a linear model controlling
for school effects whereas our approach is nearly nonparametric without
controlling for school effects. We tried to use the school dummies, but
could not, because of singularity problems due to some schools having too
few pupils; there were 80 schools.

\qquad Krueger (1999, p. 524) also shows that the positive effects of $D$
are greater for blacks, pupils with free lunch, and low-achieving pupils.
This is supported partly by Table 4, because the total effects with
endogenous $M$ are stronger for the low (0.1 and 0.3) quintiles than for the
mid (0.5 and 0.7) quintiles. However, in our analysis, the total effect
becomes stronger back again for the highest (0.9) quintile.

\section{Conclusions}

\qquad In this paper, we addressed how to decompose the total effect of an
exogenous binary treatment $D$ on an outcome $Y$, when an endogenous binary
mediator $M$ is present. The endogeneity problem was overcome with a binary
instrumental variable (IV) $Z$. We derived nonparametric \textquotedblleft
causal reduced forms (CRF's)\textquotedblright\ for $M$ and $Y$, and two
CRF's were utilized for $Y$, with one having $(1,D,Z,DZ)$ as regressors and
the other having $(1,D,M,DZ)$. The slopes of the regressors are sub-effects
that make up the total effect.

\qquad The role of $Z$ is inducing $M$ to change exogenously, but
differently from the usual endogenous treatment problem that is overcome
with an IV $Z$ where $Z$ induces an exogenous change in $D$, we required an
identification condition: \textit{the identified change that is exogenously
induced by }$Z$\textit{\ on }$M$\textit{\ should be \textquotedblleft
equivalent to\textquotedblright\ the change induced by }$D$\textit{\ on }$M$%
. This critical condition is satisfied, if all effects are constant as in
typical linear structural form (SF) models with constant effects, which
explains why this condition has been overlooked in the literature. In our
approach based on nonparametric CRF's with unrestricted effect heterogeneity
with respect to covariates $X$, we were able to discover the critical
condition because we did not impose constant effects from the outset.

\qquad Our proposed estimators are simple, as they consist of OLS to $Y$
with the regressors $(X,XD,XZ,XDZ)$, and IVE to $Y$ with the regressors $%
(X,XD,XM,XDZ)$. In both OLS and IVE, the slopes of the regressors as well as
the intercept are unknown functions of $X$, which are specified initially as
linear functions so that OLS and IVE can be easily applied. The OLS\
provides the desired total effect, and the IVE provides the direct effect;
subtracting the latter from the former then renders the indirect effect.

\qquad Going further, in case $X$ is high-dimensional, we proposed to
replace $X$ with power functions of the three-dimensional \textquotedblleft
score\textquotedblright\ $\{E(D|X),E(Z|X),E(DZ|X)\}$. Since they can be also
high-dimensional, we then proposed to replace $X$ only with power functions
of the `instrument score' $\zeta _{X}\equiv E(Z|X)$ when $D$ is randomized.
Differently from other existing effect decomposition estimators, ours are
much easier to implement, as they require only OLS and IVE despite that they
are close to being nonparametric.

\qquad We applied our estimators to a data set from the Project Star, where $%
Y$ is the grade-3 test score divided by its SD, $D$ is being in a small
class, $M$ is a binary quintile-transform of the grade-2 test score, and $Z$
is a binary quintile-transform of the grade-1 test score; we used $0.1$, $%
0.3 $, $0.5$, $0.7$ and $0.9$ quintiles. Compared with exogenous $M$,
allowing for endogenous $M$ resulted in smaller total effects. Also, whereas
the direct effect is greater than the indirect effect for all quintiles for
exogenous $M$, allowing for endogenous $M$ resulted in the indirect effect
through the grade-2 test score being greater than the direct effect for low
or high quintiles, although not for mid quintiles. This suggests stronger
indirect effects for poor or good pupils, but weaker indirect effects for
average pupils.\bigskip

\begin{center}
{\LARGE APPENDIX}
\end{center}

\textbf{A Random Effect Example for C(d) and C(e)}\medskip

\qquad Let $1[A]\equiv 1$ if $A$ holds and $0$ otherwise. For $i=1,...,N$
units, consider:%
\begin{equation*}
M_{i}^{dz}=1[0<\alpha _{1i}+\alpha _{di}d+\alpha _{zi}z+\varepsilon _{i}],\
\ \ \ \ 0\leq \alpha _{di},\alpha _{zi}.
\end{equation*}%
Then the IV-CP, TR$_{0}$-CP, and TR$_{1}$-CP hold, respectively, if the
following holds:%
\begin{eqnarray*}
&&1=M_{i}^{01}=1[0<\alpha _{1i}+\alpha _{zi}+\varepsilon _{i}],\text{\ }%
0=M_{i}^{00}=1[\alpha _{1i}+\varepsilon _{i}<0]:-\alpha _{1i}-\alpha
_{zi}<\varepsilon _{i}<-\alpha _{1i}; \\
&&1=M_{i}^{10}=1[0<\alpha _{1i}+\alpha _{di}+\varepsilon _{i}],\text{ }%
0=M_{i}^{00}=1[\alpha _{1i}+\varepsilon _{i}<0]:-\alpha _{1i}-\alpha
_{di}<\varepsilon _{i}<-\alpha _{1i}; \\
&&1=M_{i}^{11}=1[0<\alpha _{1i}+\alpha _{di}+\alpha _{zi}+\varepsilon _{i}],%
\text{ }0=M_{i}^{01}=1[\alpha _{1i}+\alpha _{zi}+\varepsilon _{i}<0]: \\
&&\ \ \ \ \ \ \ \ \ \ \ \ \ \ \ \ \ \ \ \ \ \ \ \ \ \ \ \ \ \ \ \ \ \ \ \ \
\ \ \ \ \ \ \ \ \ \ -\alpha _{1i}-\alpha _{di}-\alpha _{zi}<\varepsilon
_{i}<-\alpha _{1i}-\alpha _{zi}.
\end{eqnarray*}%
Even if $Y_{i}^{00}$ is related to $\varepsilon _{i}$ so that $M_{i}$ is
related to $Y_{i}^{00}$, if $Y_{i}^{01}-Y_{i}^{00}$ is not related to $%
\varepsilon _{i}$ because $Y_{i}^{01}$ and $Y_{i}^{00}$ contain the same
additive function of $\varepsilon _{i}$, then C(d) and C(e) hold.\bigskip

\textbf{Proof of Y-CRF1\medskip }

\qquad Since both $D$ and $M$ (but not $Z$) affect $Y$, we have%
\begin{eqnarray*}
&&Y=(1-D)(1-M)Y^{00}+(1-D)MY^{01}+D(1-M)Y^{10}+DMY^{11} \\
&&\ \ \ =Y^{00}+(Y^{10}-Y^{00})\cdot D+(Y^{01}-Y^{00})\cdot M+\Delta Y^{\pm
}\cdot DM.
\end{eqnarray*}%
Substitute (2.1) into this $Y$ equation, so that only $(D,Z,X)$ remains on
the right-hand side along with $M^{dz}$'s:%
\begin{eqnarray}
&&\ Y=Y^{00}+(Y^{10}-Y^{00})\cdot D  \notag \\
&&\ \ \ \ \ \ +(Y^{01}-Y^{00})\cdot
\{M^{00}+(M^{10}-M^{00})D+(M^{01}-M^{00})Z+\Delta M^{\pm }DZ\}  \notag \\
&&\ \ \ \ \ \ +\Delta Y^{\pm }D\cdot
\{M^{00}+(M^{10}-M^{00})D+(M^{01}-M^{00})Z+\Delta M^{\pm }DZ\}.  \notag
\end{eqnarray}%
Collect the terms with $D$, $Z$ and $DZ$: with $\Delta Y^{\pm }M^{00}+\Delta
Y^{\pm }(M^{10}-M^{00})=\Delta Y^{\pm }M^{10}$,%
\begin{eqnarray*}
&&\ Y=Y^{00}+(Y^{01}-Y^{00})M^{00} \\
&&\ +\{Y^{10}-Y^{00}+(Y^{01}-Y^{00})(M^{10}-M^{00})+\Delta Y^{\pm
}M^{10}\}D+(Y^{01}-Y^{00})(M^{01}-M^{00})Z \\
&&\ +\{(Y^{01}-Y^{00})\Delta M^{\pm }+\Delta Y^{\pm }(M^{01}-M^{00})+\Delta
Y^{\pm }\Delta M^{\pm }\}DZ.
\end{eqnarray*}

\qquad Take $E(\cdot |D,Z,X)$ on this $Y$ equation to invoke the $(D,Z)$%
-exogeneity in C(a):%
\begin{eqnarray*}
&&E(Y|D,Z,X)=E\{Y^{00}+(Y^{01}-Y^{00})M^{00}|X\} \\
&&+E\{Y^{10}-Y^{00}+(Y^{01}-Y^{00})(M^{10}-M^{00})+\Delta Y^{\pm
}M^{10}|X\}\cdot D \\
&&+E\{(Y^{01}-Y^{00})(M^{01}-M^{00})|X\}\cdot Z \\
&&+E\{(Y^{01}-Y^{00})\Delta M^{\pm }+\Delta Y^{\pm }(M^{01}-M^{00})+\Delta
Y^{\pm }\Delta M^{\pm }|X\}\cdot DZ.
\end{eqnarray*}%
The slope of $DZ$ can be further simplified to $(Y^{01}-Y^{00})\Delta M^{\pm
}+\Delta Y^{\pm }(M^{11}-M^{10})$. Using this and defining $U_{1}\equiv
Y-E(Y|D,Z,X)$ renders Y-CRF1.\bigskip \newpage \textbf{Proof of
Y-CRF2\medskip }

\qquad Adding and subtracting a few terms, rewrite the regression function
of Y-CRF1 $E(Y|D,Z,X)=\beta _{0}(X)+\beta _{d}(X)D+\beta _{z}(X)Z+\beta
_{dz}(X)DZ$:%
\begin{eqnarray*}
&&\ E(Y|D,Z,X)=\{\beta _{0}(X)-\beta _{m}(X)\alpha _{0}(X)\}\ +\ \{\beta
_{d}(X)-\beta _{m}(X)\alpha _{d}(X)\}D+\beta _{m}(X)M \\
&&\ +\beta _{m}(X)\alpha _{0}(X)+\beta _{m}(X)\alpha _{d}(X)D+\beta
_{m}(X)\alpha _{z}(X)Z+\beta _{m}(X)\alpha _{dz}(X)DZ-\beta _{m}(X)M \\
&&\ +\{\beta _{z}(X)-\beta _{m}(X)\alpha _{z}(X)\}Z+\{\beta _{dz}(X)-\beta
_{m}(X)\alpha _{dz}(X)\}DZ.
\end{eqnarray*}%
The five terms in the middle with $\beta _{m}(X)$ can be written as%
\begin{equation*}
\beta _{m}(X)\{\alpha _{0}(X)+\alpha _{d}(X)D+\alpha _{z}(X)Z+\alpha
_{dz}(X)DZ-M\}=\beta _{m}(X)(-U_{0});
\end{equation*}%
$E\{\beta _{m}(X)U_{0}|D,Z,X\}=0$ holds by construction. Also, as is
discussed in Remark 2, $\{\beta _{z}(X)-\beta _{m}(X)\alpha _{z}(X)\}Z=0$
due to C(d). Hence, we obtain%
\begin{eqnarray*}
&&\ E(Y|D,Z,X)=\{\beta _{0}(X)-\beta _{m}(X)\alpha _{0}(X)\}\ +\ \{\beta
_{d}(X)-\beta _{m}(X)\alpha _{d}(X)\}D+\beta _{m}(X)M \\
&&\ +\{\beta _{dz}(X)-\beta _{m}(X)\alpha _{dz}(X)\}DZ\ -\beta _{m}(X)U_{0}.
\end{eqnarray*}

Finally, the definition of $U_{2}$ renders Y-CRF2.\bigskip

\textbf{Proof of Remark 4\medskip }

\qquad Note $\alpha _{d}(X)+\alpha _{dz}(X)=E(M^{10}-M^{00}|X)+E(\Delta
M^{\pm }|X)=E(M^{11}-M^{01}|X)$. Using this and recalling (2.4), the slope
of $D$ in Y-CRF2 when $Z=1$ is%
\begin{eqnarray*}
&&\ \beta _{d}(X)+\beta _{dz}(X)\ -\ \beta _{m}(X)\{\alpha _{d}(X)+\alpha
_{dz}(X)\} \\
&&\ =E\{Y^{10}-Y^{00}+(Y^{01}-Y^{00})(M^{11}-M^{01})+\Delta Y^{\pm
}M^{11}|X\} \\
&&\ \ \ \ \ -E(Y^{01}-Y^{00}|M^{10}-M^{00}=1,X)\cdot E(M^{11}-M^{01}|X).
\end{eqnarray*}%
In the second term here, invoke C(e) so that $M^{10}-M^{00}$ can be replaced
with $M^{11}-M^{01}$. Then, the second term becomes $%
E\{(Y^{01}-Y^{00})(M^{11}-M^{01})|X\}$, which cancels out the middle
indirect effect in the first term. Hence, we obtain%
\begin{equation*}
\beta _{d}(X)+\beta _{dz}(X)-\beta _{m}(X)\{\alpha _{d}(X)+\alpha
_{dz}(X)\}=E(Y^{10}-Y^{00}+\Delta Y^{\pm }M^{11}|X).
\end{equation*}

\newpage

\begin{center}
{\LARGE REFERENCES}
\end{center}

\qquad Angrist, J.D., 2001, Estimation of limited dependent variable models
with dummy endogenous regressors, Journal of Business and Economic
Statistics 19, 2-28.

\qquad Bellani, L. and M. Bia, 2019, The long-run effect of childhood
poverty and the mediating role of education, Journal of the Royal
Statistical Society (Series A) 182, 37-68.

\qquad Burgess, S., R.M. Daniel, A.S. Butterworth and S.G. Thompson, 2015,
Network Mendelian randomization: using genetic variants as instrumental
variables to investigate mediation in causal pathways, International Journal
of Epidemiology 44, 484-495.

\qquad Chen, S.H., Y.C. Chen and J.T. Liu, 2019, The impact of family
composition on educational achievement, Journal of Human Resources 54,
122-170.

\qquad Choi, J.Y., G. Lee and M.J. Lee, 2023, Endogenous treatment effect
for any response conditional on control propensity score, Statistics and
Probability Letters 196, 109747.

\qquad Choi, J.Y. and M.J. Lee, 2018, Regression discontinuity with multiple
running variables allowing partial effects, Political Analysis 26, 258-274.

\qquad Choi, J.Y. and M.J. Lee, 2023a, Overlap weight and propensity score
residual for heterogeneous effects:\ a review with extensions, Journal of
Statistical Planning and Inference 222, 22-37.

\qquad Choi, J.Y. and M.J. Lee, 2023b, Complier and monotonicity for fuzzy
multi-score regression discontinuity with partial effects, Economics Letters
228, 111169.

\qquad Fr\"{o}lich M. and M. Huber, 2017, Direct and indirect treatment
effects-causal chains and mediation analysis with instrumental variables,
Journal of the Royal Statistical Society (Series B) 79, 1645-1666.

\qquad Huber, M., M. Lechner and A. Strittmatter, 2018, Direct and indirect
effects of training vouchers for the unemployed, Journal of the Royal
Statistical Society (Series A) 181, 441-463.

\qquad Imai, K., L. Keele and T. Yamamoto, 2010, Identification, inference,
and sensitivity analysis for causal mediation effects, Statistical Science
25, 51-71.

\qquad Imai, K., D. Tingley and T. Yamamoto, 2013, Experimental designs for
identifying causal mechanisms, Journal of the Royal Statistical Society
(Series A) 176, 5-32.

\qquad Imbens, G.W. and J.D. Angrist, 1994, Identification and estimation of
local average treatment effects, Econometrica 62, 467-475.

\qquad Joffe, M.M., D. Small, T.T. Have, S. Brunelli and H.I. Feldman, 2008,
Extended instrumental variables estimation for overall effects,
International Journal of Biostatistics 4 (1), Article 4.

\qquad Kim, B.R., 2025, Estimating spillover effects in the presence of
isolated nodes, Spatial Economic Analysis, 1-15.

\qquad Kim, B.R. and M.J. Lee, 2024, Instrument-residual estimator for
multi-valued instruments under full monotonicity, Statistics and Probability
Letters 213, 110187.

\qquad Krueger, A.B., 1999, Experimental estimates of education production
functions, Quarterly Journal of Economics 114, 497-532.

\qquad Lee, G., J.Y. Choi and M.J. Lee, 2023, Minimally capturing
heterogeneous complier effect of endogenous treatment for any outcome
variable, Journal of Causal Inference 11 (1), 20220036.

\qquad Lee, M.J. 2018, Simple least squares estimator for treatment effects
using propensity score residuals, Biometrika 105, 149-164.

\qquad Lee, M.J., 2021, Instrument residual estimator for any response
variable with endogenous binary treatment, Journal of the Royal Statistical
Society (Series B) 83, 612-635.

\qquad Lee, M.J., 2024, Direct, indirect and interaction effects based on
principal stratification with a binary mediator, Journal of Causal Inference
12, 20230025\textit{.}

\qquad Lee, M.J. and C. Han, 2024, Ordinary least squares and
instrumental-variables estimators for any outcome and heterogeneity, Stata
Journal 24, 72-92.

\qquad Lee, M.J., G. Lee and J.Y. Choi, 2025,\ Linear probability model
revisited: why it works and how it should be specified, Sociological Methods
\& Research 54, 173-186.

\qquad Lee, M.J. and S.H. Lee, 2022, Review and comparison of treatment
effect estimators using propensity and prognostic scores, International
Journal of Biostatistics 18, 357-380.

\qquad Mao, H. and L. Li, 2020, Flexible regression approach to propensity
score analysis and its relationship with matching and weighting, Statistics
in Medicine 39, 2017-2034.

\qquad Mattei, A. and F. Mealli, 2011, Augmented designs to assess principal
strata direct effects, Journal of the Royal Statistical Society (Series B)
73, 729-752.

\qquad MacKinnon, D.P., A.J. Fairchild and M.S. Fritz, 2007, Mediation
analysis, Annual Review of Psychology 58, 593-614.

\qquad Miles, C.H., 2023, On the causal interpretation of randomised
interventional indirect effects, Journal of the Royal Statistical Society
(Series B) 85, 1154-1172.

\qquad Nguyen, T.Q., I. Schmid and E.A. Stuart, 2021, Clarifying causal
mediation analysis for the applied researcher: defining effects based on
what we want to learn, Psychological Methods 26, 255-271.

\qquad Pearl, J., 2001, Direct and indirect effects, in Proceedings of the
Seventeenth Conference on Uncertainty in Artificial Intelligence, San
Francisco, CA, Morgan Kaufman, 411-420.

\qquad Pearl, J., 2009, Causality, 2nd ed., Cambridge University Press.

\qquad Preacher, K.J. 2015, Advances in mediation analysis: a survey and
synthesis of new developments, Annual Review of Psychology 66, 825-852.

\qquad Robins, J.M., 2003, Semantics of causal DAG models and the
identification of direct and indirect effects, In Highly Structured
Stochastic Systems, edited by P.J. Green, N.L. Hjort and S. Richardson,
70-81, Oxford University Press, Oxford.

\qquad Rosenbaum, P.R. and D.B. Rubin, 1983, The central role of the
propensity score in observational studies for causal effects, Biometrika 70,
41-55.

\qquad Rudolph, K.E., N. Williams, and I. Diaz, 2024, Using instrumental
variables to address unmeasured confounding in causal mediation analysis,
Biometrics 80, ujad037.

\qquad Stock, J.H. and M.W. Watson, 2007, Introduction to Econometrics, 2nd
ed., Addison Wesley.

\qquad TenHave, T.R. and M.M. Joffe, 2012, A review of causal estimation of
effects in mediation analyses, Statistical Methods in Medical Research 21,
77-107.

\qquad VanderWeele, T.J., 2013, A three-way decomposition of a total effect
into direct, indirect, and interactive effects, Epidemiology 24, 224-232.

\qquad VanderWeele, T.J., 2014, A unification of mediation and interaction:
a four-way decomposition, Epidemiology 25, 749-761.

\qquad VanderWeele, T.J., 2015, Explanation in causal inference: methods for
mediation and interaction, Oxford University Press.

\end{document}